\documentclass[ reprint,
superscriptaddress,
 amsmath,amssymb,
 aps, floatfix
]{revtex4-2}

\usepackage{graphicx}
\usepackage{dcolumn}
\usepackage{bm}

\usepackage{amsmath}
\usepackage{url}
\usepackage{array,tabularx}
\usepackage{graphics} 
\usepackage{epsfig} 
\usepackage{mathptmx} 
\usepackage{times} 
\usepackage{amssymb}  
\usepackage{xfrac}
\usepackage{comment}
\usepackage{xcolor}
\usepackage{epstopdf}
\usepackage{gensymb}

\usepackage{mathtools}
\usepackage[export]{adjustbox}

\newcommand\x{3.9}

\date{}

\let\OLDthebibliography\thebibliography
\renewcommand\thebibliography[1]{
  \OLDthebibliography{#1}
  \setlength{\parskip}{0.5pt}
  \setlength{\itemsep}{0.5pt plus 0.3ex}
}

\begin{document}


\title{Klein-like tunneling of sound via negative index metamaterials} 
\author{Lea Sirota}\email{leabeilkin@tauex.tau.ac.il}

\affiliation{School of Mechanical Engineering, Tel Aviv University, Tel-Aviv 69978, Israel}

\date{\today}



\begin{abstract}
Klein tunneling 
is a counterintuitive quantum-mechanical phenomenon, predicting perfect transmission of relativistic particles through higher energy barriers.
This phenomenon was shown to be supported at normal incidence in graphene
due to pseudospin conservation.
Here I show that Klein tunneling analogue can occur in classical systems, and remarkably, not relying on mimicking graphene's spinor wavefunction structure.
Instead, the mechanism requires
a particular form of constitutive parameters of the penetrated medium, yielding transmission properties identical to the quantum tunneling in graphene. 
I demonstrate this result by simulating tunneling of sound in a two-dimensional acoustic metamaterial.
More strikingly, I show that by introducing a certain form of anisotropy, the tunneling can be made unimpeded for any incidence angle, while keeping most of its original Klein dispersion properties. 
This phenomenon may be denoted by the omnidirectional Klein-like tunneling. The new tunneling mechanism and its omnidirectional variant may be useful for applications requiring lossless and direction-independent transmission of classical waves.
\end{abstract}

\maketitle

\section{Introduction}

The idea to guide classical waves by mimicking quantum-mechanical wave phenomena has received a major interest in recent years.
This is enabled due to the striking analogy between the electronic band-structure of solids and the frequency dispersion of classical systems \cite{franz2013topological}.
For example, a great deal of attention was devoted to mimicking quantum topological phenomena \cite{thouless1982quantized,haldane1988model,kane2005quantum,bernevig2006quantum} in acoustic and elastic media, realizing it using architectured materials or metamaterials.
The topological properties of the band-structure were exploited to achieve unique functionalities that are uncommon for sound and vibration, 
such as beam-like narrow waves, which are immune to backscattering from corners, bents, and structural defects \cite{khanikaev2015topologically,mousavi2015topologically,zhang2017topological,vila2017observation,chaunsali2018subwavelength,hofmann2019chiral,brandenbourger2019non,sirota2020non,sirota2020feedbackA,sirota2020real,scheibner2020non,rosa2020dynamics,darabi2020experimental}. 

However, an entire class of quantum-mechanical phenomena related to tunneling remains considerably under-explored for classical waveguiding. 
These phenomena include Klein tunneling of relativistic particles 
\cite{klein1929reflexion,katsnelson2006chiral,huard2007transport,stander2009evidence,allain2011klein,robinson2012klein}, tunneling of particles across the event horizon of black holes \cite{hawking1975particle}, tunneling of electron pairs through superconducting junctions \cite{voss1981macroscopic}, and more. 
The common property of these effects, which constitutes the essence of tunneling, is an unusual and counterintuitive ability of particles to cross gaps, barriers or interfaces, despite this crossing being forbidden in a sense by dynamical or energetic considerations. 
Translating this exciting property into the classical realm holds the potential to substantially advance waveguiding capabilities in classical systems.
In this work the focus is on Klein tunneling.

\section{The original quantum effect}

\begin{figure}[tb] 
\begin{center}
\begin{tabular}{l l}
   \textbf{a}  &  \textbf{b} \\
   \includegraphics[width=3.6 cm, valign=c]{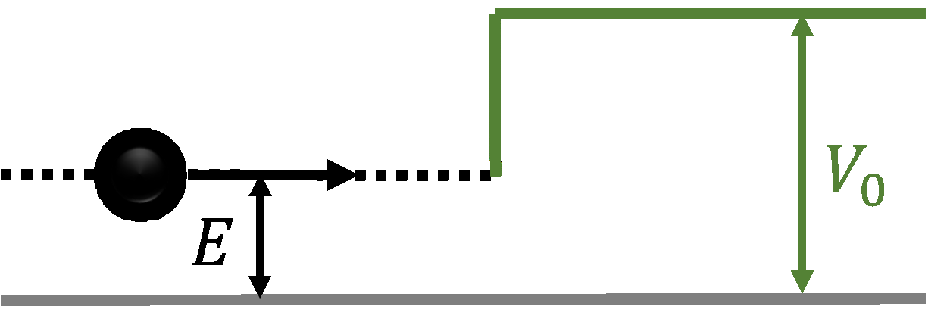} & \includegraphics[width=3.6 cm, valign=c]{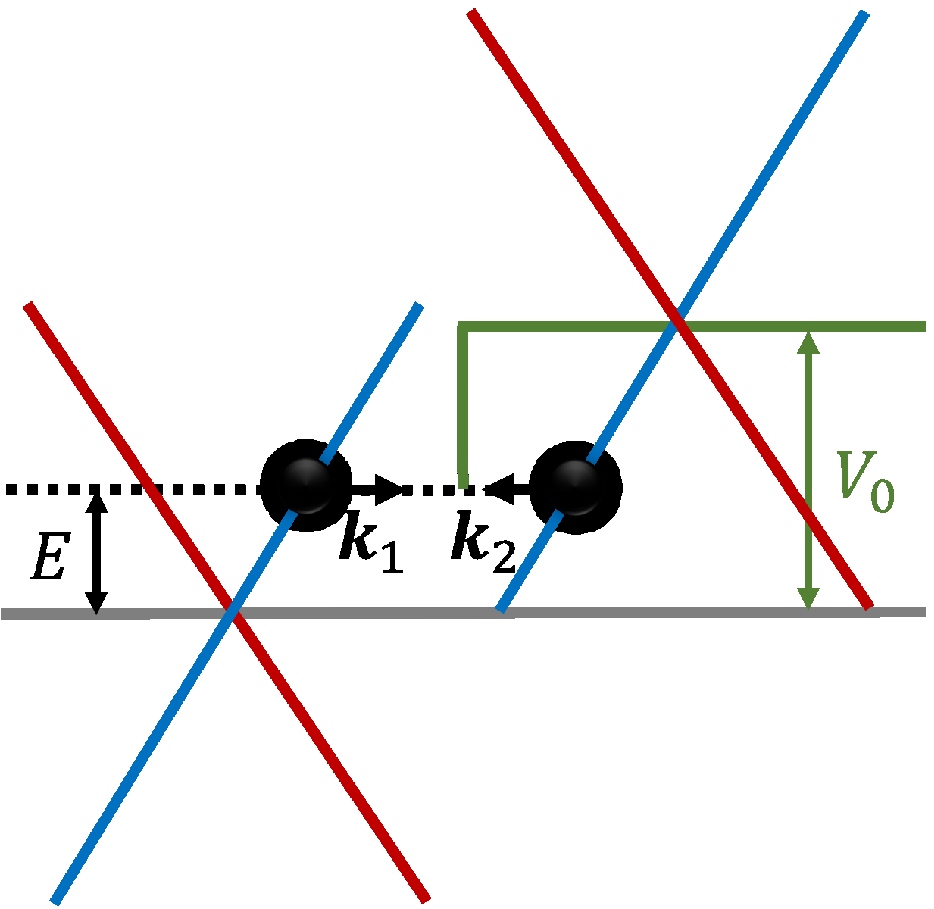} \\
  \textbf{c}   &  \textbf{d} \\
  \includegraphics[width=3.6 cm, valign=c]{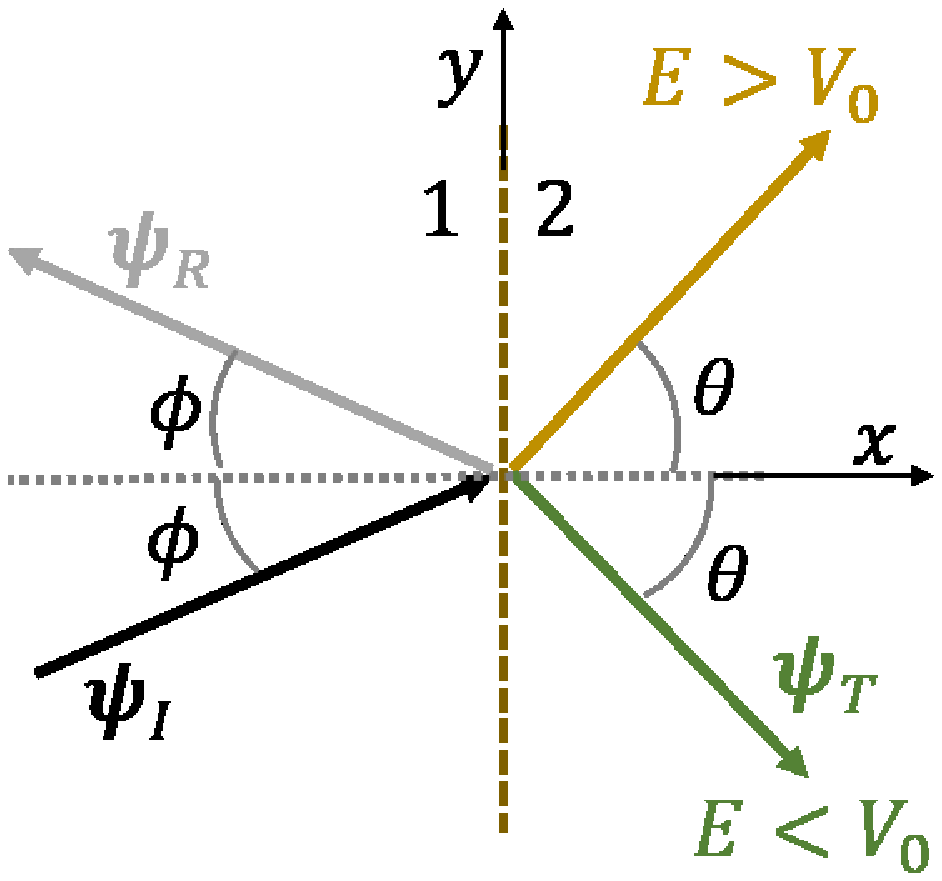}   &  \includegraphics[width=3.6 cm, valign=c]{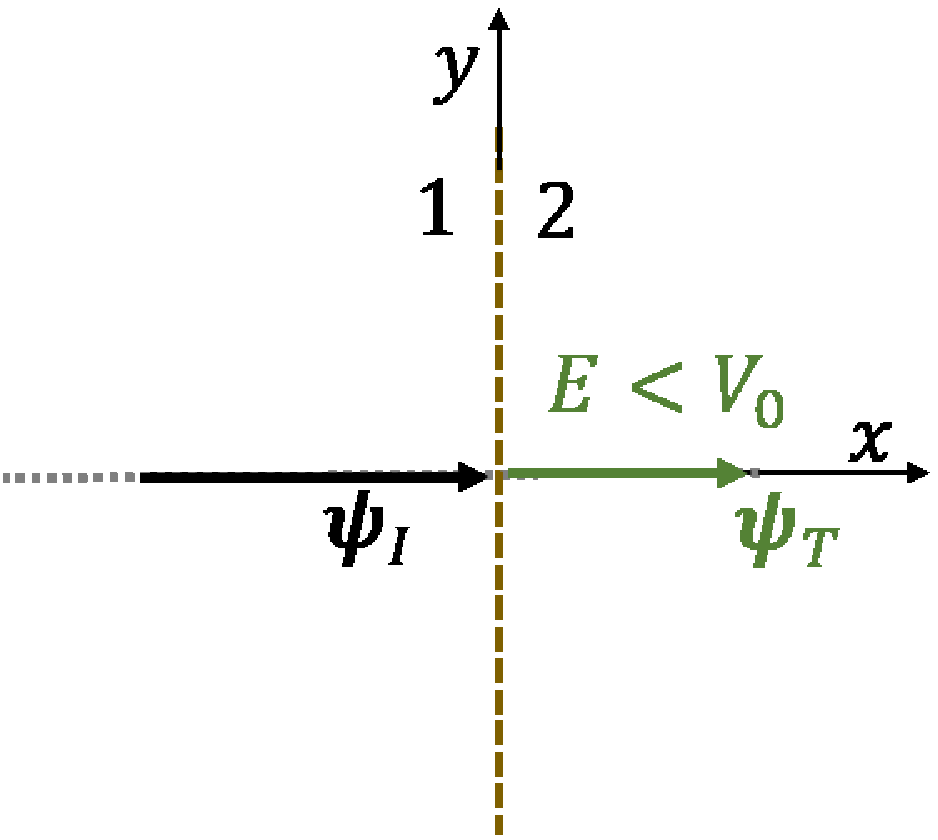}
\end{tabular}
\end{center}
 \caption{(a) The quantum Klein paradox schematic. (b) Klein tunneling in graphene illustrating particle transition to the lower Dirac cone with addition of potential $V_0>E$. 
 (c) Wave $\psi_I$ (energy $E$, momentum $k_1$) incident in graphene at $\phi>0$, between domains $1$ and $2$ that differ by the energy $V_0$. The transmitted wave $\psi_T$ (energy $E$, momentum $k_2$) is at angle $\theta$, negative for $E<V_0$ (the tunneling case) and positive for $E>V_0$. Reflected wave $\psi_R$ exists. (d) Klein tunneling in graphene at normal incidence $\phi=0$. The transmission is reflectionless for any $V_0$.
 }
\label{Fig_1}
\end{figure}

Quantum tunneling described by the Klein paradox \cite{klein1929reflexion} is a phenomenon, in which relativistic particles unimpededly cross a potential barrier regardless of its height and width, Fig. \ref{Fig_1}(a). 
The fact that this crossing has a unity transmission probability when the barrier energy $V_0$ is higher than the particle energy $E$
is counterintuitive, as one would expect the transmission probability to decay with an increasing barrier height, as in the non-relativistic scenario. 

A similar effect was predicted 
\cite{katsnelson2006chiral}, observed \cite{huard2007transport,stander2009evidence}, and analyzed \cite{allain2011klein,robinson2012klein} for Dirac electrons in graphene between two domains that differ by a constant electrostatic potential $V(x)=V_0$. 
The underlying principle for tunneling in graphene was shown to originate from the two-component structure of its wavefunctions, which resembles Dirac spinors, and features Dirac-like cone dispersion, Fig.  \ref{Fig_1}(b).
At the transmission to the higher potential side the electron of energy $E$ and momentum $k_1$ is shifted to the lower band with the same energy but a different momentum $k_2$, keeping its velocity direction but flipping its momentum direction. 

In Figs. \ref{Fig_1}(c),(d) a step potential of height $V_0$ and infinite width is considered. The domains with $V(x)=0$ at $-\infty<x<0$ and with $V(x)=V_0$ at $0\leq x<\infty$ are labeled by $1$ and $2$.
For a wave $\psi_I$ incident from domain $1$, and a wave $\psi_T$ transmitted to domain $2$, 
Fig. \ref{Fig_1}(c), 
the incidence and transmission angles $\phi$ and $\theta$ constitute the phases between the two components of the respective wavefunctions, and are associated with sublattice pseudospin.
For $\phi>0$ a reflected wave $\psi_R$ exists.
Momentum equity in the $\textbf{y}$ direction in both domains, together with wavefunction continuity at the domain's interface referring to pseudospin conservation \cite{allain2011klein}, yields
\begin{subequations}  \label{eq:Klein_eqs}
\begin{align}
   & \sin\phi=\frac{E-V_0}{E}\sin\theta,   \label{eq:Klein_eqs_Snell} \\
   & \begin{cases}
   1+R=T, \\
   (1-R)\cos\phi=T\cos\theta,
   \end{cases}   \label{eq:Klein_eqs_Fresnel}
\end{align}
\end{subequations}
where $R$ and $T$ are reflection and transmission amplitudes. The relation in Eq. \eqref{eq:Klein_eqs_Snell} implies that for a given $\phi$, the relative `heights' of $E$ and $V_0$ are translated to the sign of $\theta$ (and $k_2$), which is positive for $E>V_0$ and negative for the tunneling case $E<V_0$.
The tunneling effect is manifested in Eq. \eqref{eq:Klein_eqs_Fresnel} at normal incidence ($\phi=0$), where the transmission becomes unimpeded irrespective of $E$ and $V_0$, implying $R=0$ and $T=1$, as depicted in Fig. \ref{Fig_1}(d) for $E<V_0$. 

The exotic properties of Klein tunneling inspired the search for analogies in other systems \cite{bahat2010klein,ni2018spin,jiang2020direct}, but were exclusively based on mimicking graphene or graphene-like lattices. 
Next I demonstrate that a tunneling effect with properties identical to Eq. \eqref{eq:Klein_eqs_Snell}-\eqref{eq:Klein_eqs_Fresnel} can occur in inherently classical systems without a restriction to the particular graphene's wavefunction structure and dispersion. 

\section{A non-spinor classical analogy}

\subsection{Effective medium model}

To this end I consider the system in Fig. \ref{Fig_1}(c),(d) to represent a continuous acoustic medium defined by pressure field $p(x,y,t)$ and flow velocity field $\textbf{v}(x,y,t)$.
Domain $1$ is a uniform acoustic medium of mass density $m_0$ and bulk modulus $b_0$.
Domain $2$ is a complex medium, described by dynamical mass density $m_0\widetilde{M}(\omega)$ and bulk modulus $b_0\widetilde{B}(\omega)$. $\omega$ is the sound wave frequency. 
The constitutive parameters $\widetilde{M}(\omega)$ and $\widetilde{B}(\omega)$ play a crucial role in reproducing Klein-like tunneling in this system.
Assuming longitudinal wave propagation \cite{bruneau2013fundamentals} and time-harmonic dependence $p_j(x,y,t)=P_j(x,y)e^{-i\omega t}$, $\textbf{v}_j(x,y,t)=\textbf{V}_j(x,y)e^{-i\omega t}$, $j=1,2$ indicating domain number, this system is governed by
\begin{subequations}  \label{eq:Medium_eq}
\begin{align}
    \nabla P_j(x,y)&=i\omega m_0\widetilde{M}_j(\omega)\textbf{V}_j(x,y),   \label{eq:Medium_eq_M} \\ i\omega P_j(x,y)&=b_0\widetilde{B}_j(\omega) \nabla\cdot \textbf{V}_j(x,y).   \label{eq:Medium_eq_B}
\end{align}
\end{subequations}
Here $\widetilde{M}_1(\omega)=\widetilde{B}_1(\omega)=1$, $\widetilde{M}_2(\omega)=\widetilde{M}(\omega)$ and $\widetilde{B}_2(\omega)=\widetilde{B}(\omega)$. 
I now consider a pressure wave of amplitude $P_I$ incident from domain $1$ at angle $\phi$, a reflected wave $P_R$, and a wave $P_T$ transmitted to domain $2$ at angle $\theta$, which respectively stand for $\psi_I$, $\psi_R$ and $\psi_T$ in Fig. \ref{Fig_1}(c).
Employing horizontal stratification, continuity of pressure, and continuity of normal flow velocity along the domain's interface (derivation details appear in Appendix A), gives
\begin{subequations}  \label{eq:Klein_like_eqs}
\begin{align}
   & \sin\phi=\left(\widetilde{M}(\omega)\widetilde{B}^{-1}(\omega)\right)^{1/2}\sin\theta,   \label{eq:Klein_like_eqs_Snell} \\
   & \begin{cases}
   1+R=T, \\
   (1-R)\cos\phi=\left(\widetilde{M}(\omega)\widetilde{B}(\omega)\right)^{-1/2}T\cos\theta.
   \end{cases}   \label{eq:Klein_like_eqs_Fresnel}
\end{align}
\end{subequations}
These classical concepts of wave propagation between media, Snell's law of refraction in Eq. \eqref{eq:Klein_like_eqs_Snell}, and Fresnel's reflection and transmission coefficients $R$ and $T$ in Eq. \eqref{eq:Klein_like_eqs_Fresnel}, are strikingly similar to the quantum tunneling properties in Eq. \eqref{eq:Klein_eqs_Snell} and \eqref{eq:Klein_eqs_Fresnel}.
Using the mapping $E \leftrightarrow \omega^2$,
the matching of Eq. \eqref{eq:Klein_like_eqs_Snell}-\eqref{eq:Klein_like_eqs_Fresnel} to Eq. \eqref{eq:Klein_eqs_Snell}-\eqref{eq:Klein_eqs_Fresnel} gives
\begin{subequations}  \label{eq:Klein_like_conds}
\begin{align}
  \left(\widetilde{M}(\omega)\widetilde{B}^{-1}(\omega)\right)^{1/2}&=\frac{\omega^2-V_0}{\omega^2},    \label{eq:MB_explicit} \\
  \left(\widetilde{M}(\omega)\widetilde{B}(\omega)\right)^{1/2}&=1.  \label{eq:MB_impedance} 
\end{align}
\end{subequations}
This determines the mass density and bulk modulus as
\begin{equation}
    \widetilde{M}(\omega)=\frac{\omega^2-V_0}{\omega^2} \quad , \quad \widetilde{B}(\omega)=\frac{\omega^2}{\omega^2-V_0}.   \label{eq:MB}
\end{equation}
The particular combination of the parameters in Eq. \eqref{eq:MB}
creates acoustic tunneling with properties identical to the tunneling of electrons in graphene, although the underlying continuous fields physics in Eq. \eqref{eq:Medium_eq_M}-\eqref{eq:Medium_eq_B} is fundamentally different from the quantum Dirac physics. 
In Eq. \eqref{eq:MB_explicit}, 
$(\widetilde{M}(\omega)\widetilde{B}^{-1}(\omega))^{1/2}=k_2/k_1$, with $k_1=\omega/c$, $c=(b_0/m_0)^{1/2}$, is the ratio of domain $1$ and $2$ wavenumbers, resulting in
\begin{equation}  \label{eq:k1k2}
    \frac{k_2}{k_1}=\frac{\omega^2-V_0}{\omega^2} \quad , \quad \theta=\tan^{-1}\frac{k_{1y}}{k_{2x}}.
\end{equation}
Both the wavenumber ratio $k_2/k_1$ and the transmission angle $\theta$ in Eq. \eqref{eq:k1k2} perfectly coincide with corresponding values of the quantum Klein tunneling in graphene \cite{katsnelson2006chiral,allain2011klein,robinson2012klein}.
In Eq. \eqref{eq:MB_impedance},
$(\widetilde{M}(\omega)\widetilde{B}(\omega))^{1/2}=z_2/z_1$, with $z_1=(m_0b_0)^{1/2}$, is the respective ratio of the specific acoustic impedance \cite{pierce2019acoustics}.  
Eq. \eqref{eq:MB_impedance} indicates that the impedance of domains $2$ and $1$ is matched at all frequencies. This implies that a normally-incident acoustic wave for any $\omega^2/V_0$ (and an obliquely-incident wave for the particular case $\omega^2/V_0=\frac{1}{2}$) will penetrate domain $2$ completely free of backscattering, i.e. with $R=0$, similarly to the quantum tunneling. For all other values of $\omega^2/V_0$, $R\neq 0$ at oblique incidence.
However, impedance matching alone is not enough for the analogy; the particular dispersion of Eq. \eqref{eq:k1k2} is required in domain $2$. 

Here, $k_2$ in Eq. \eqref{eq:k1k2}, and $\widetilde{M}(\omega)$ and $\widetilde{B}(\omega)$ in Eq. \eqref{eq:MB} are positive for $\omega^2>V_0$, and negative for $\omega^2<V_0$. 
The notion of a negative wavenumber, resulting from simultaneously negative constitutive parameters, is a celebrated concept in the research of wave propagation in electromagnetic and acoustic systems.
It indicates antiparallel phase and group velocities, leading to extraordinary phenomena unavailable in natural materials \cite{engheta2006metamaterials,caloz2008crlh,seo2012acoustic,cummer2016controlling,baz2010active,sirota2019tunable}. 
In fact, the expressions in Eq. \eqref{eq:MB} coincide with the so-called matched Drude model \cite{engheta2006metamaterials} or left-handed electric networks \cite{caloz2008crlh},
in the electromagnetic terminology. 
In this section I showed that Eq. \eqref{eq:MB} constitutes an exact classical analogue of Klein tunneling. Next I propose its realization using an acoustic metamaterial. 
\begin{figure}[tb] 
\begin{center}
\begin{tabular}{c}
  \textbf{a}   \\
  \includegraphics[width=7.2 cm, valign=c]{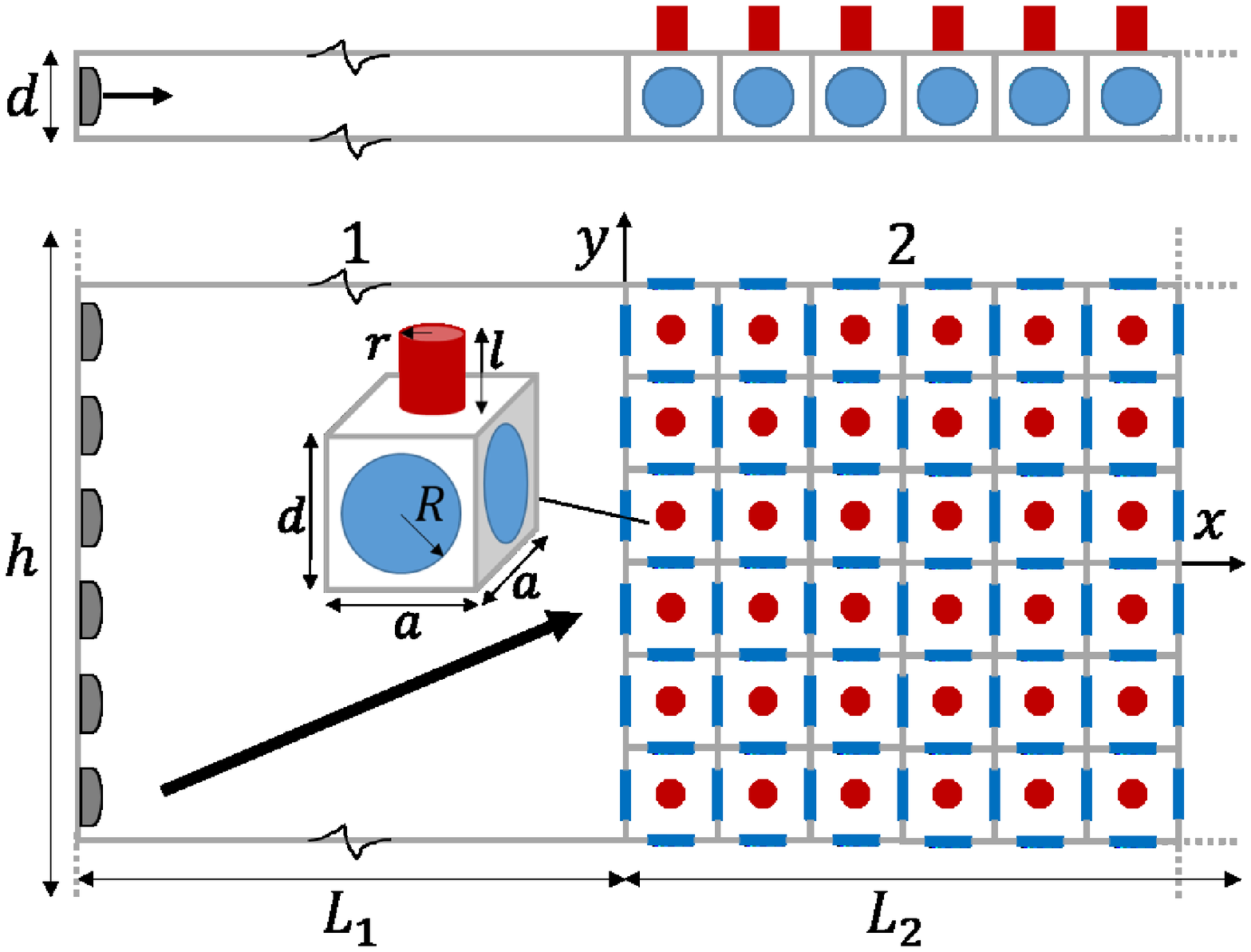}
  \end{tabular}
  \begin{tabular}{c c}
      \textbf{b}  &  \textbf{c} \\
  \includegraphics[height=4.2 cm, valign=c]{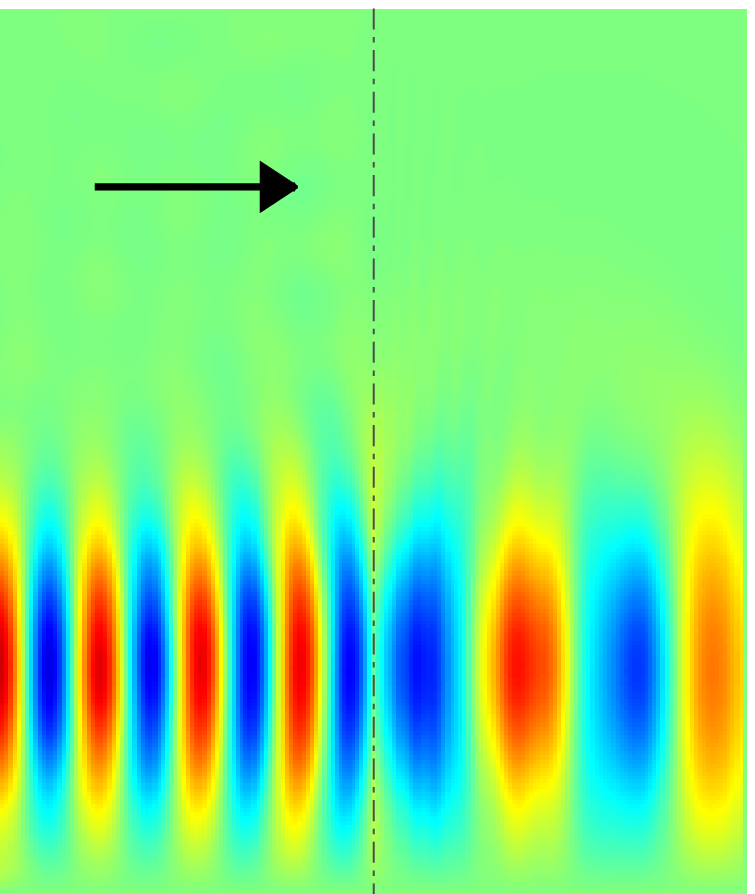} &
  \includegraphics[height=4.2 cm, valign=c]{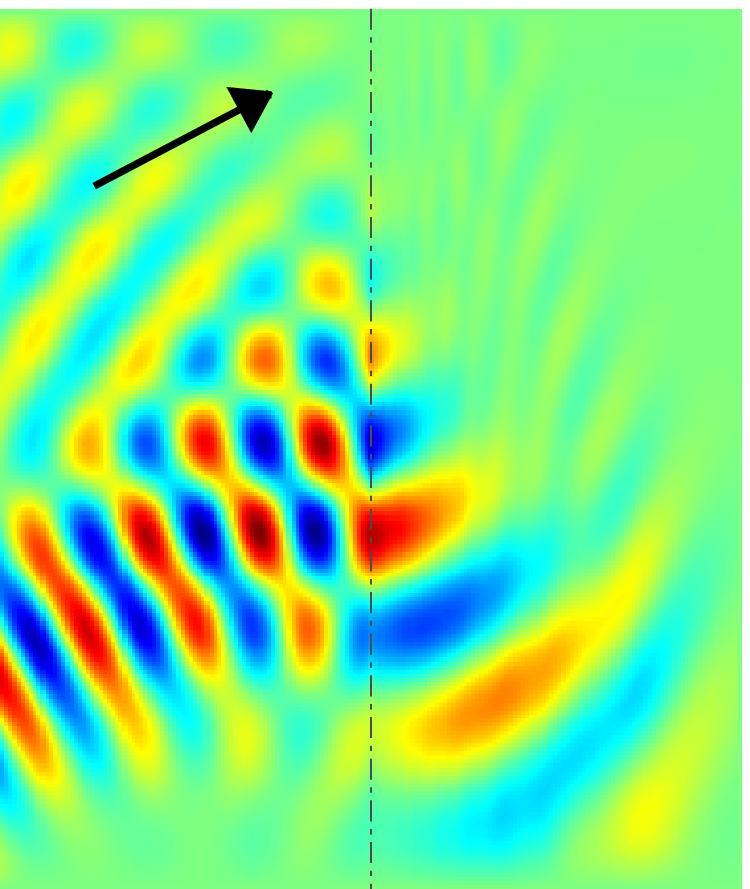}
\end{tabular}
\begin{tabular}{c}
\\
     \includegraphics[width=3.8  cm, valign=c]{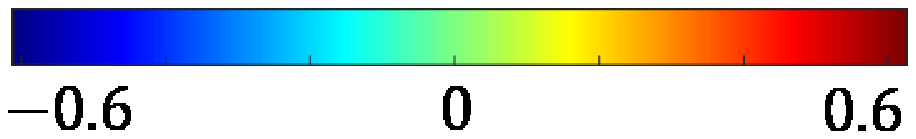}
\end{tabular}
\end{center}
 \caption{(a) Acoustic metamaterial schematic mimicking Klein tunneling. Side and top view. (b)-(c) The time domain responses of a $h=180$ $[cm]$, $L_1=75$ $[cm]$ and $L_2=75$ $[cm]$ homogenized system in air ($m_0=1.2$ $[kg/m^3]$, $b_0=1.42\cdot 10^5$ $[Pa]$) with phase and group velocities $v_{p1}=v_{g1}=c=344$ $[m/s]$. The sources generate a beam of frequency $\omega_0=1.72$ $[kHz]$, implying wavelength $\lambda_1=20$ $[cm]$ and wavenumber $k_1=31.4$ $[r/m]$ in domain 1. Setting $\gamma_0=\omega_0^2/V_0=2/3$, gives $V_0=1.75\cdot 10^8$ $[r^2/s^2]$, $k_2=-k_1/2$, $\lambda_2=2\lambda_1$, and $v_{p2}=-v_{g2}=-2c$ in domain 2, indicating negative refraction. (b) Normal incidence $\phi=0^o$. No reflection occurs. (c) Oblique incidence $\phi=28^o$. Substantial reflection occurs.}
\label{Fig_2}
\end{figure}

\subsection{Acoustic metamaterial realization}

The proposed metamaterial is illustrated in Fig. \ref{Fig_2}(a).
Domain $1$ is a waveguide of area $L_1\times h$, consisting of two rigid parallel plates, gapped by a distance $d$.
Domain $2$, of area $L_2\times h$, is a matrix of $a\times a \times d$ cuboids, Fig. \ref{Fig_2}(a) inset, with elastic membranes (blue circles) of radius $R$ and stiffness $B_m$ ($\frac{1}{2}B_m$ for a unit cell) mounted in the walls, and an open side branch cavity resonator of length $l$ and radius $r$ \cite{pierce2019acoustics} (red cylinder) at the top. 
The external walls are sealed, with an array of acoustic actuators (grey circles) at the left wall, producing source waves (black arrow). 

The membranes create an effective mass density of $\widetilde{M}(\omega)=(\omega^2-\omega_m^2)/\omega^2$, $\omega_m^2=B_m/(m_0a^2d)$, whereas the resonator creates an effective bulk modulus of $\widetilde{B}(\omega)=\omega^2/(\omega^2-\omega_b^2)$, $\omega_b^2=\pi r^2c^2/(a^2dl)$ (derivation details appear in Appendix B).
To satisfy Eq. \eqref{eq:MB}, $\omega_m^2=\omega_b^2$ must hold, yielding $B_m=\pi r^2b_0/l$ and $V_0=\pi r^2c^2/(a^2dl)$. 
The graphene potential $V_0$ thus translates into a function of the metamaterial's constitutive parameters and geometry, and unlike the quantum system does not represent any physical addition. It indicates the threshold between a double-positive and a double-negative index acoustic medium.
The particular value of $V_0$ depends on the desired ratio $\gamma=\omega^2/V_0$.
The wavelength in domain $2$, $\lambda_2=2\pi/k_2$, is determined from Eq. \eqref{eq:k1k2}.
For $a\ll \lambda_2$, the collective unit cell dynamics turns the metamaterial into an effectively continuous material with properties determined by Eq. \eqref{eq:MB}. 

The tunneling is demonstrated in dynamical simulations of a homogenized metamaterial of overall size $h=180$ $[cm]$, $L_1=75$ $[cm]$ and $L_2=75$ $[cm]$. The medium is air with $m_0=1.2$ $[kg/m^3]$, $b_0=1.42\cdot 10^5$ $[Pa]$, implying the phase (and group) velocity $v_{p1}=c=344$ $[m/s]$. The sources generate a beam of frequency $\omega_0=1.72$ $[kHz]$, leading to the wavelength $\lambda_1=20$ $[cm]$ (or wavenumber $k_1=31.4$ $[r/m]$). To obtain $\omega_0^2<V_0$, as required for tunneling, we set, e.g. $\gamma_0=\omega_0^2/V_0=2/3$. This results in $V_0=1.75\cdot 10^8$ $[r^2/s^2]$, and by Eq. \eqref{eq:k_12} in $k_2=-k_1/2$ and $\lambda_2=2\lambda_1$. 
The pressure fields, obtained by the finite difference time domain (FDTD) method (simulation construction details appear in Appendix D), are plotted for two cases.
In the first, Fig. \ref{Fig_2}(b), the source beam incidence is normal, $\phi=0^o$, resulting in a perfectly unimpeded tunneling.
In the second, Fig. \ref{Fig_2}(c), the beam is incident at $\phi=28^o$, resulting in tunneling at a negative angle $\theta\approx -70^o$, and a partial reflection. In both cases the refraction is negative with the phase and group velocities $v_{p2}=-v_{g2}=-2c$. The simulated transmission angles, wavelengths and wave velocities are in full agreement with the theoretical expectations from Eqs. \eqref{eq:Klein_like_conds}-\eqref{eq:MB}.

\section{Omnidirectional Klein-like tunneling}

\subsection{Anisotropic medium design}  \label{Anisoderiv}

\begin{figure}[tb]
     \centering
     \includegraphics[width=7.2 cm, valign=c]{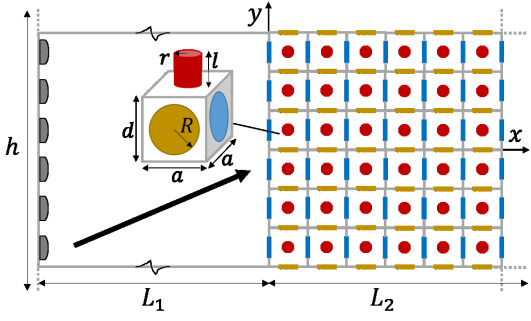}
     \caption{Schematic of an acoustic metamaterial supporting omnidirectional Klein-like  tunneling.}
     \label{fig:aniso_scheme}
 \end{figure}

It would be exceptionally interesting to discover conditions for which the Klein-like tunneling defined by Eq. \eqref{eq:Klein_like_conds} becomes unimpeded regardless of the incidence angle, for any $\omega^2/V_0$.
This could be useful for applications that require navigating detection beams of arbitrary incidence angles and frequencies around an object without backscattering (acoustic camouflaging, for example).
Since Eq. \eqref{eq:MB_explicit} and \eqref{eq:MB_impedance} uniquely determine the metamaterial parameters, an additional degree of freedom in the design is required.
This can be obtained by introducing anisotropy \cite{chen2005retrieval,akl2012technique} to the effective mass density, with $\widetilde{M}_x(\omega)$ in the $\textbf{x}$ direction and $\widetilde{M}_y(\omega)\neq\widetilde{M}_x(\omega)$ in the $\textbf{y}$ direction. 
The system is then described by Eq. \eqref{eq:Medium_eq} with two distinct equations in Eq. \eqref{eq:Medium_eq_M}.
A possible realization in an acoustic metamaterial is illustrated in Fig. \ref{fig:aniso_scheme}, which is similar to the one in Fig. \ref{Fig_2}(a), but with $\textbf{y}$ axis membranes (yellow bars) of a different stiffness than the $\textbf{x}$ axis membranes. 

Continuity of pressure along the domain's interface gives $1+R=T$, similarly to Eq. \eqref{eq:Klein_eqs_Fresnel}. The distinction from the original effect is manifested in the continuity of normal flow velocity.
Instead of matching it with the conditions in Eq. \eqref{eq:Klein_eqs}, the substitution $k_{2x}=k_{1x}\widetilde{M}_x(\omega)$
should be used (details are given in Appendix C).
Adding horizontal stratification $k_{1y}=k_{2y}$ leads to $\widetilde{M}_x(\omega)\widetilde{B}(\omega)=1$ and $\widetilde{B}(\omega)\widetilde{M}_y^{-1}(\omega)=1$. To relate to the Klein dispersion of Eq. \eqref{eq:MB_explicit},
the substitution $\widetilde{M}_x(\omega)=\widetilde{M}(\omega)$ is necessary, with $\widetilde{M}(\omega)$ and $\widetilde{B}(\omega)$ defined in Eq. \eqref{eq:MB}. This results in 
\begin{subequations}  \label{eq:Klein_aniso_eqs}
\begin{align}
   & \tan\phi=\frac{\omega^2-V_0}{\omega^2}\tan\theta,   \label{eq:Klein_aniso_eqs_Snell} \\
   & R=0, \quad T=1.   \label{eq:Klein_aniso_eqs_Fresnel}
\end{align}
\end{subequations}
The relations in Eq. \eqref{eq:Klein_aniso_eqs_Snell}-\eqref{eq:Klein_aniso_eqs_Fresnel} resemble Eq. \eqref{eq:Klein_like_eqs_Snell}-\eqref{eq:Klein_like_eqs_Fresnel}, yet are essentially different. Eq. \eqref{eq:Klein_aniso_eqs_Snell}, which may be considered as a modified Snell's law of refraction, indicates that there is no critical angle for any $\gamma=\omega^2/V_0$ and $\phi$, yet the refractive angle $\theta$ is positive for $\gamma>1$ and negative for $\gamma<1$, as in the original effect (Fig. \ref{contour} in Appendix C). 
Eq. \eqref{eq:Klein_aniso_eqs_Fresnel} indicates unimpeded transmission to domain $2$ for any $\gamma$ and $\phi$. I denote this effect by omnidirectional Klein-like tunneling.

 %
The parameter $\widetilde{M}_y(\omega)$ can be of any form, provided the overall system is dynamically stable, where the condition $\widetilde{B}(\omega)\widetilde{M}_y^{-1}(\omega)=1$ is
applied to a specific working frequency $\omega_0$, i.e. $\widetilde{M}_y(\omega_0)=\omega_0^2/(\omega_0^2-V_0)$. For example,
\begin{equation}   \label{eq:My_aniso}
    \widetilde{M}_y(\omega)=\frac{\omega^2-\alpha}{\omega^2} \quad , \quad \alpha=\frac{\gamma_0}{1-\gamma_0}V_0,
\end{equation}
where $\gamma_0=\omega_0^2/V_0<1$.
For $\gamma_0=1/2$, $\widetilde{M}_y(\omega)$ in Eq. \eqref{eq:My_aniso} retrieves $\widetilde{M}(\omega)$ of Eq. \eqref{eq:MB_explicit}.
The anisotropic medium defined by Eq. \eqref{eq:My_aniso} thus supports unimpeded transmission for any incidence angle $\phi$ and a particular frequency $\omega=\omega_0$. To support a different frequency, the parameter $\alpha$ in $\widetilde{M}_y(\omega)$ needs to be adjusted accordingly.
The constitutive parameters determine the medium's dispersion relation, which then takes the form 
\begin{equation}   \label{eq:k1k2_aniso}
    \frac{k_{2x}}{k_{1x}}=\frac{\omega^2-V_0}{\omega^2} \quad , \quad \theta=\tan^{-1}\frac{k_{1y}}{k_{2x}},
\end{equation}
with $k_{1y}=k_{2y}=(\omega/c)\sin\phi$.
This relation captures the underlying mechanism of the omnidirectional tunneling.
In fact, Eq. \eqref{eq:k1k2_aniso} is the $\textbf{x}$ axis projection of the original Klein dispersion in Eq. \eqref{eq:k1k2}, indicating that the omnidirectional $k_2$ has the same Klein-like dispersion as in the angle-dependent case, just scaled by the positive constant $\cos\phi/\cos\theta$. 
Contrary to the quantum graphene, for which the dispersion at the vicinity of Dirac points consists of two touching cones both in domains $1$ and $2$, with a constant shift of $V_0$ in domain $2$, Fig. \ref{Fig_1}(b), the situation for the omnidirectional acoustic analogue is quite different, as discussed next.

\subsection{Tilted cones dispersion and dynamical response}

\begin{figure*}[tb] 
\begin{center}
  \begin{tabular}{c c c c}
  \textbf{a}    & \textbf{b}  & \textbf{c}   & \textbf{d}  \\
  Domain 1 & Domain 2, $\gamma_0=\frac{2}{3}$  & Domain 2, $\gamma_0=\frac{1}{2}$  & Domain 2, $\gamma_0=\frac{1}{3}$ \\
    \includegraphics[height=5.0 cm, valign=c]{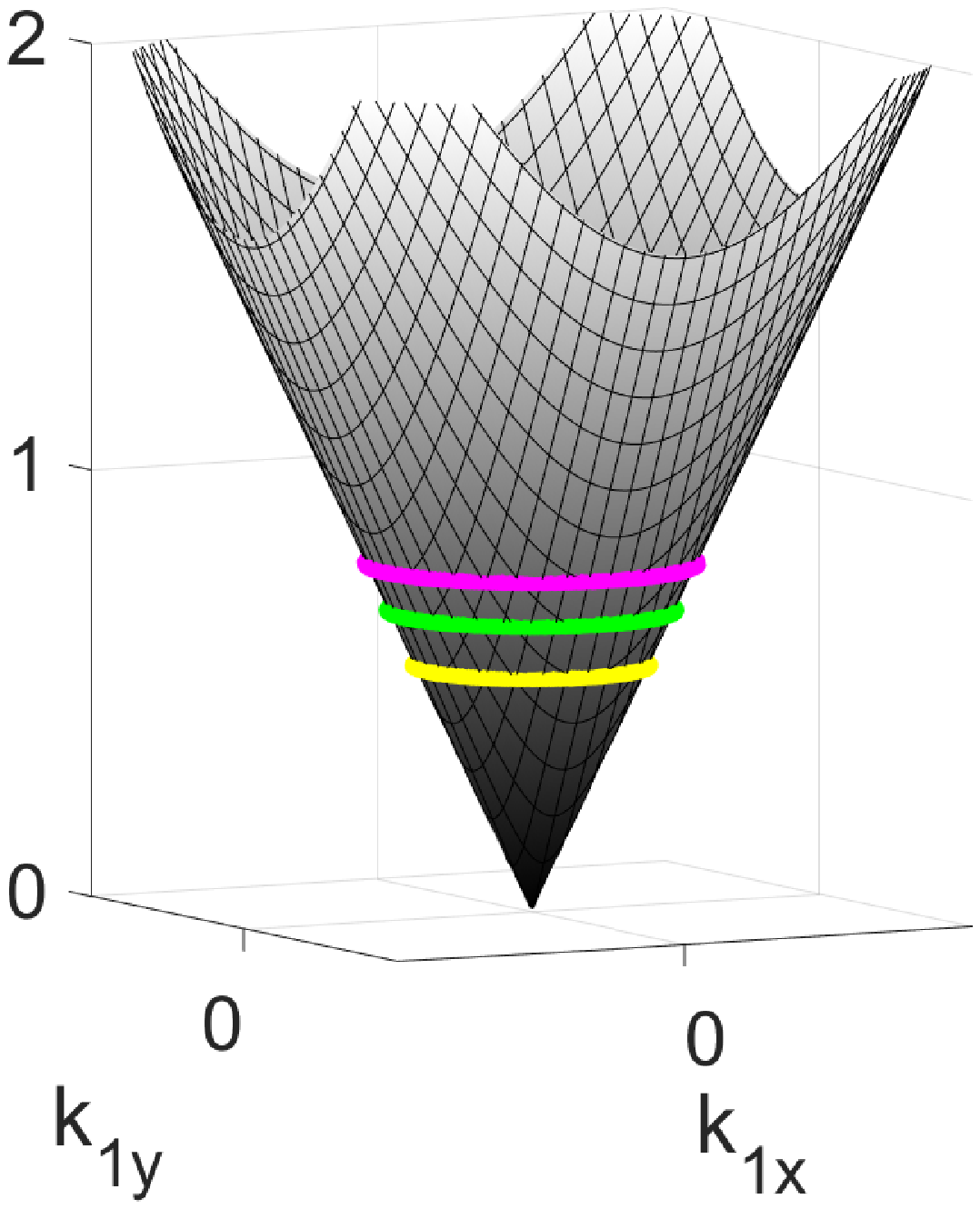} & 
  \includegraphics[height=5.0 cm, valign=c]{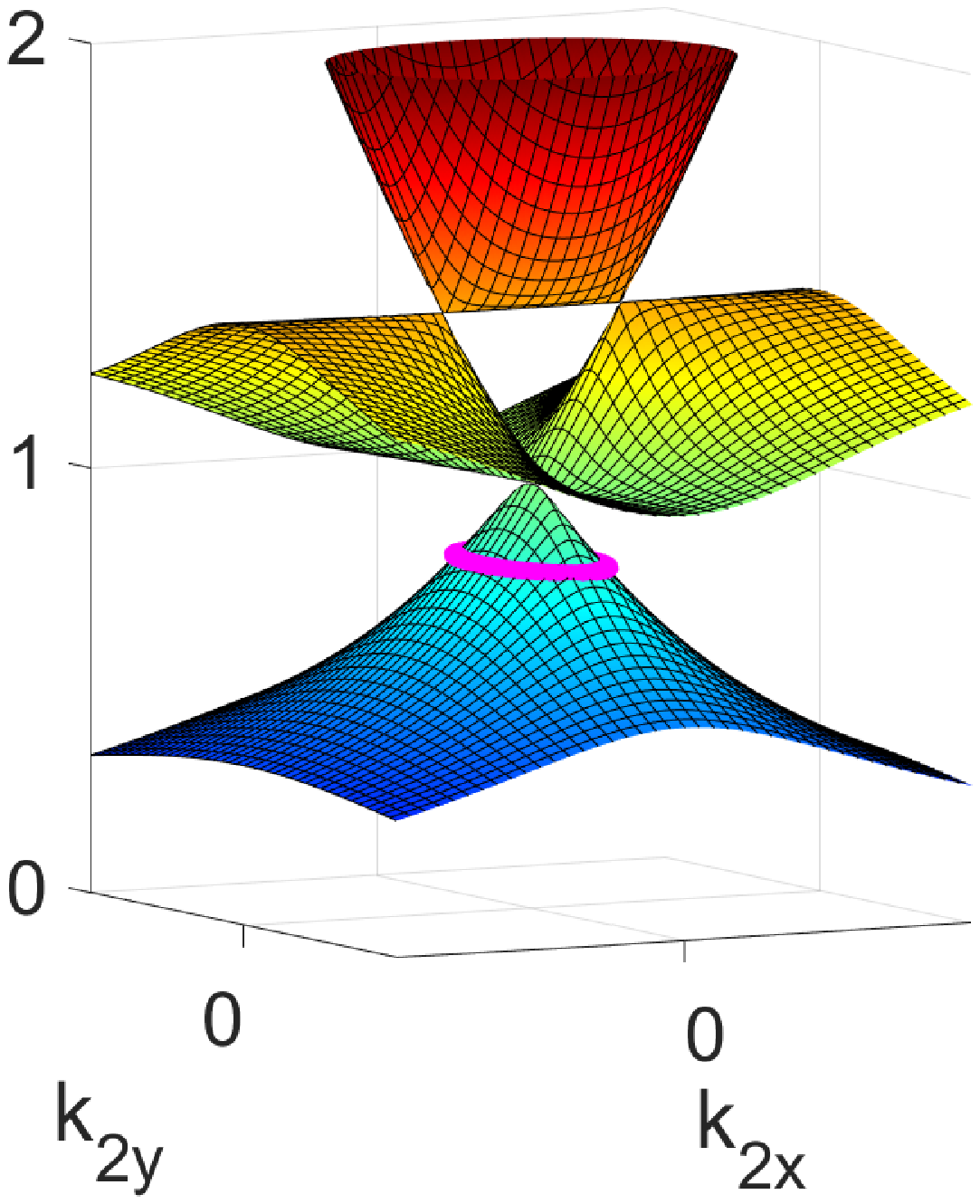} & \includegraphics[height=5.0 cm, valign=c]{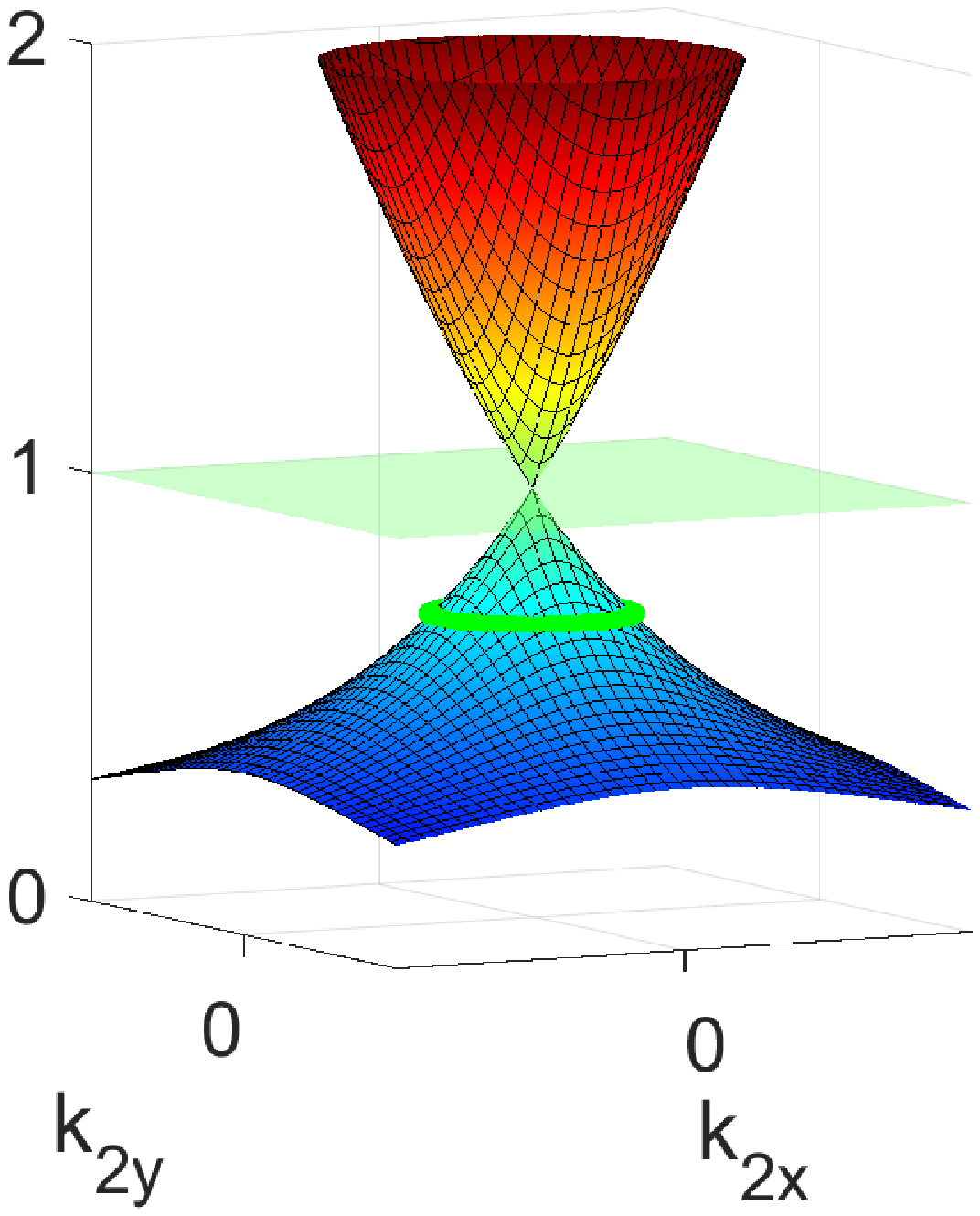} & \includegraphics[height=5.0 cm, valign=c]{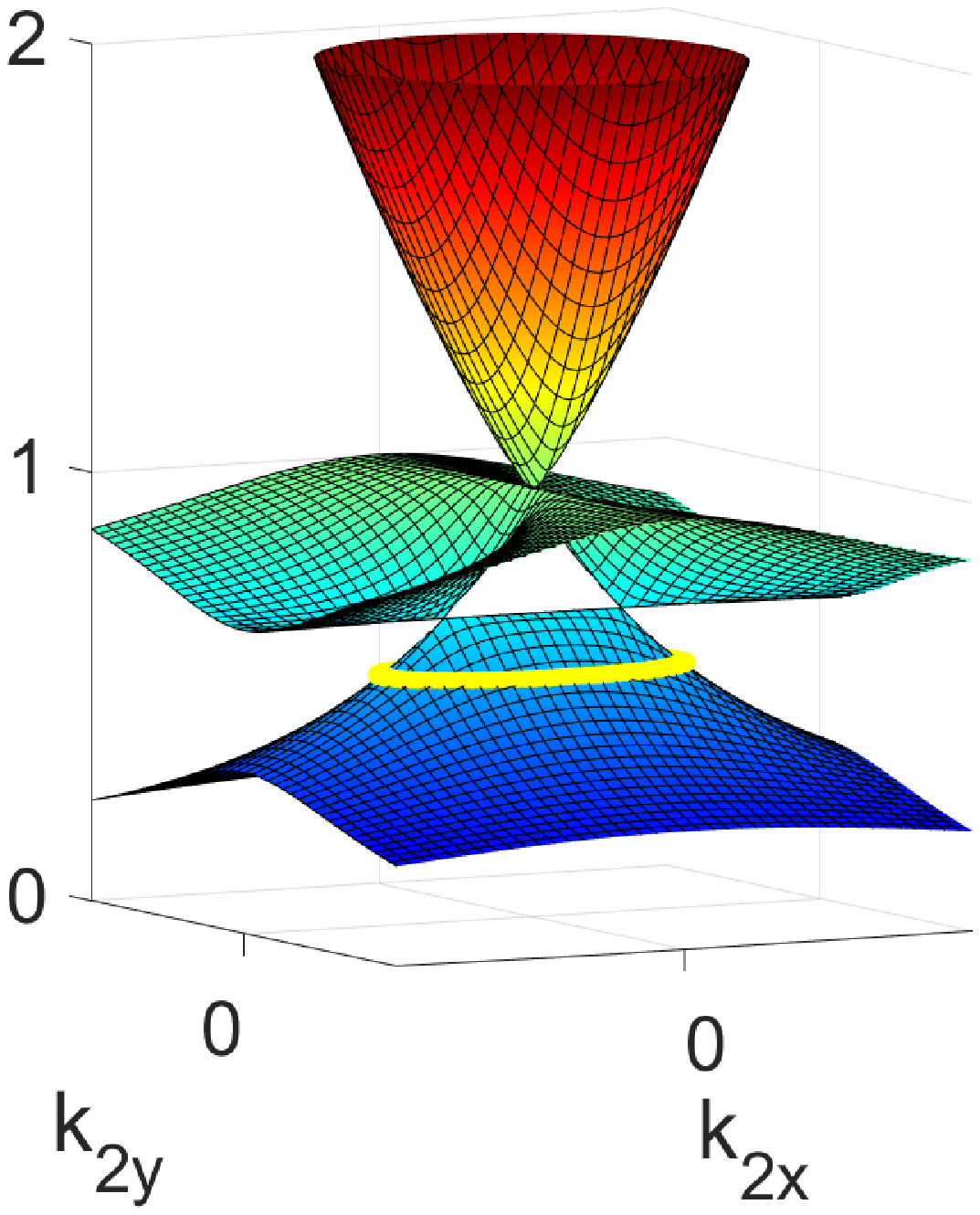}  \\
       & \textbf{e}  & \textbf{f}  & \textbf{g} \\
    \includegraphics[height=3.7  cm, valign=c]{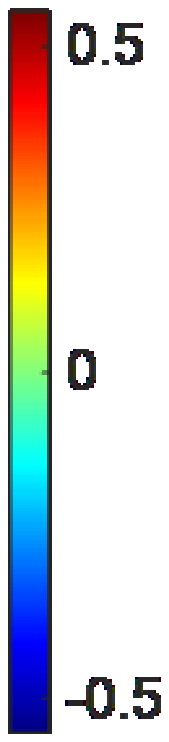} &
  \includegraphics[height=4.2 cm, valign=c]{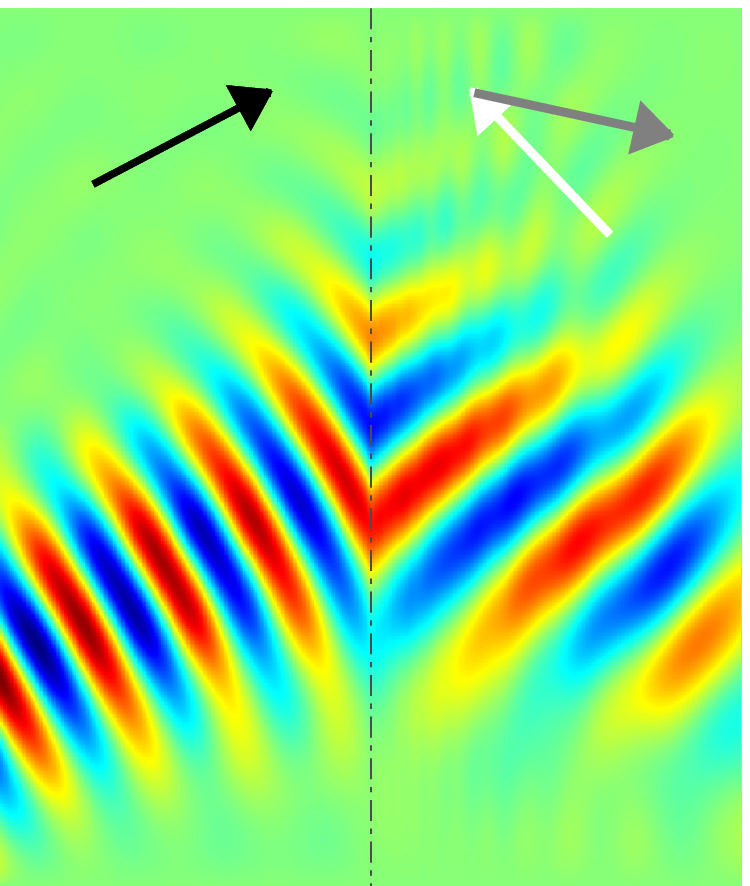} & \includegraphics[height=4.2  cm, valign=c]{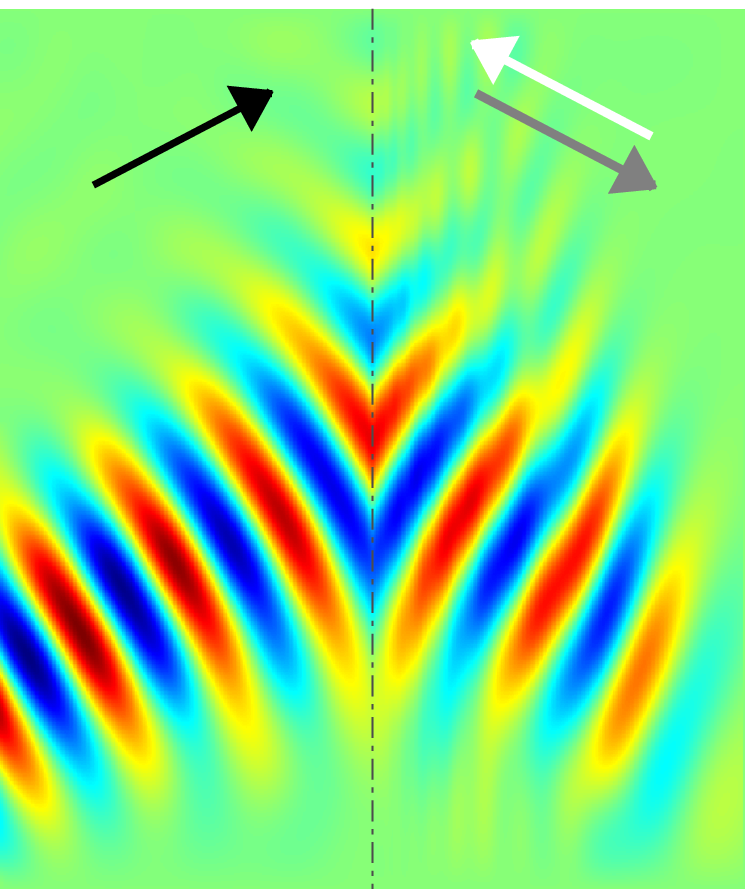} & \includegraphics[height=4.2  cm, valign=c]{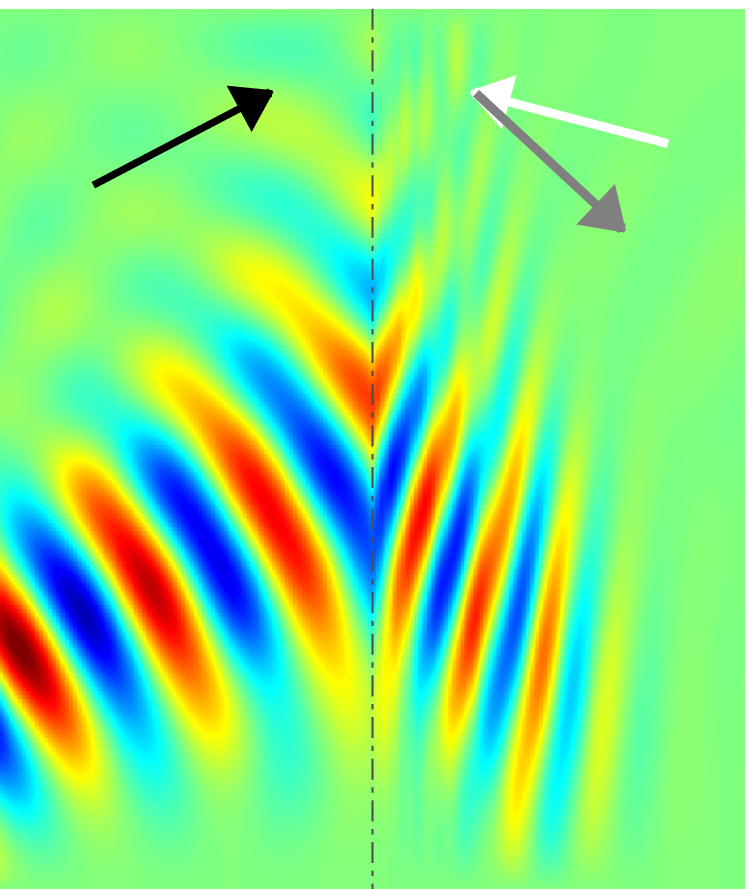}
\end{tabular}
\end{center}
 \caption{Omnidirectional acoustic Klein-like tunneling. (a) Dispersion profile in domain $1$ of the metamaterial in Fig. \ref{fig:aniso_scheme}. (b)-(d) Dispersion profiles in (homogenized) domain $2$ for $\gamma_0=2/3$, $\gamma_0=1/2$, and $\gamma_0=1/3$.
 The tunneling to domain $2$ is expressed by transition to the lower cone, with the corresponding working region highlighted by purple, green and yellow rings. 
 (e)-(g) The time domain pressure field response to a source at $\phi=28^o$ for $\gamma_0=2/3$, $\gamma_0=1/2$, and $\gamma_0=1/3$. 
 The resonators and $\textbf{x}$ direction membranes are the same as in the isotropic case, featuring $V_0=1.75\cdot 10^8$ $[r^2/s^2]$ potential. The group and phase velocities directions are indicated by grey and white arrows (amplitudes not indicated), respectively.
For $\gamma_0=2/3$, the source is given by $\omega_0=1.72$ $[kHz]$, $\lambda_1=20$ $[cm]$, as in the isotropic simulation of Fig. \ref{Fig_2}. 
The phase and group velocity directions are $\theta=-46.76^o$ and $\theta_g=12.4^o$, with $\lambda_2=2\lambda_1$ and $|v_{p2}|=0.5|v_{p1}|$. 
For $\gamma_0=1/2$, $\omega_0=1.49$ $[kHz]$, and $\lambda_1=23.1$ $[cm]$. The phase and group velocities align at $\theta=\mp 28^o$, as expected, with $\lambda_2=\lambda_1$ and $|v_{p2}|=|v_{p1}|$.
For $\gamma_0=1/3$, $\omega_0=1.22$ $[kHz]$ and $\lambda_1=28.3$ $[cm]$. The velocities directions are interchanged, with $\theta=-14.9^o$, $\theta_g=43.3^o$, $\lambda_2<\lambda_1$, and $|v_{p2}|=2|v_{p1}|$. 
}
\label{Fig_3}
\end{figure*}

The dispersion of the anisotropic effective medium designed in Sec. \ref{Anisoderiv} is depicted in Fig. \ref{Fig_3}(a)-(d).
In domain $1$ the dispersion is a single circular cone, Fig. \ref{Fig_3}(a).
With the introduction of potential, expressed through the anisotropic constitutive parameters of domain $2$, this cone transforms into three surfaces, the form of which depends on $\gamma_0$, as depicted in Figs. \ref{Fig_3}(b),(c),(d) for $\gamma_0=2/3$, $1/2$ and $1/3$. 
The different values of $\gamma_0$ indicate different working frequencies, $\omega_0^2=\gamma_0V_0$, highlighted by a purple, green and yellow curve, respectively.
$V_0$ is kept constant at the value set in Eq. \eqref{eq:MB}. 

For $1/2<\gamma_0<1$ and $0<\gamma_0<1/2$, Figs. \ref{Fig_3}(b) and (d), the middle surface consists of two tilted cones of a hyperbolic isofrequency cross-section.
The upper and the lower cones are of an elliptic cross-section, respectively forming tilted Dirac-like cones with the middle surface \cite{kawarabayashi2012generalization}.
%
At $\gamma_0=1/2$ the middle surface degenerates, as captured by the transparent sheet in Fig. \ref{Fig_3}(c), and is no longer a part of the solution. The top and bottom surfaces become regular circular cones, touching at the origin. 

For any $\gamma_0$, the lower surface corresponds to the wave transition to domain $2$, similarly to the electron transition from the upper to the lower cone in quantum graphene.
For $1/2<\gamma_0<1$ ($0<\gamma_0<1/2$) the major axis of the elliptic lower cone is $k_{2x}$ ($k_{2y}$).
This polarization flipping, as illustrated by Fig. \ref{contour} in Appendix C, corresponds to the interplay of the tunnelled wave group and phase velocity directions, respectively given by the lower cone dispersion gradient in Figs.  \ref{Fig_3}(b)-(d) and the transmission angle $\theta$ in Eq. \eqref{eq:Klein_aniso_eqs_Snell}.

The omnidirectional tunneling is demonstrated in the dynamical FDTD responses of the anisotropic homogenized effective medium for three different working frequencies, $\gamma_0=2/3$, $\gamma_0=1/2$, and $\gamma_0=1/3$, respectively depicted in Figs. \ref{Fig_3}(e)-(g) (simulation details and responses of an actual discrete structure appear in Appendix C, Fig. \ref{Fig_5}).
The simulated systems feature the same bulk modulus $\widetilde{B}(\omega)$ and $\textbf{x}$ direction mass density $\widetilde{M}_x(\omega)=\widetilde{M}(\omega)$ as in the isotropic simulation in Fig. \ref{Fig_2} (the actual metamaterials will comprise the same resonators geometry and $\textbf{x}$ direction membranes stiffness as in the isotropic case), keeping $V_0=1.75\cdot 10^8$ $[r^2/s^2]$ in all the three simulations.
To accommodate the different working frequencies implied by the different $\gamma_0$, the $\textbf{y}$ direction mass density $\widetilde{M}_y(\omega)$ is changed according to \eqref{eq:My_aniso}. 
The incidence angle is $\phi=28^o$ in all the three cases. Simulations for other incidence angles are given in Fig. \ref{Fig_4} of Appendix C.

%
The resulting acoustic pressure fields demonstrate a complete transmission from medium $1$ to medium $2$ (up to minor numerical reflections).
The phase and group velocity directions polarization is respectively illustrated by the grey and white arrows at angles $\theta$ and $\theta_g$. 
For $\gamma_0=2/3$, the source is of frequency $\omega_0=1.72$ $[kHz]$ and wavelength $\lambda_1=20$ $[cm]$, as in the isotropic simulation of Fig.  \ref{Fig_2}. The transmission wavelength and phase velocity amplitude are given by $\lambda_2=2\lambda_1$ and $|v_{p2}|=0.5|v_{p1}|$, with the angles $\theta=-46.76^o$ and $\theta_g=12.4^o$. 
For $\gamma_0=1/2$, the source signal is of $\omega_0=1.49$ $[kHz]$ and $\lambda_1=23.1$ $[cm]$. The phase and group velocities align at $\theta=\mp 28^o$, as expected, with $\lambda_2=\lambda_1$ and $|v_{p2}|=|v_{p1}|$.
For $\gamma_0=1/3$, $\omega_0=1.22$ $[kHz]$ and $\lambda_1=28.3$ $[cm]$, resulting in $\lambda_2<\lambda_1$, and $|v_{p2}|=2|v_{p1}|$. The velocities directions are interchanged, with $\theta=-14.9^o$, $\theta_g=43.3^o$. 
The unimpeded negative refraction, the wavelengths and the wave velocities of the responses in Figs. \ref{Fig_3}(e)-(g) are in exact accordance with the omnidirectional tunneling properties of Eqs. \eqref{eq:Klein_aniso_eqs}-\eqref{eq:k1k2_aniso}.

\section{Conclusion}

This work provided an exact analogue of the quantum Klein tunneling phenomenon in an inherently classical acoustic medium, without mimicking graphene's spinors, but by tailoring the constitutive parameters according to Eq. \eqref{eq:MB}. Realization of these parameters in the acoustic metamaterial of Fig. \ref{Fig_2}(a) was suggested. 
Furthermore, the anisotropic design of Fig. \ref{fig:aniso_scheme}, with the tuning parameter in Eq. \eqref{eq:My_aniso}, enabled the sound to tunnel independently of incidence angle and frequency-potential ratio, obeying the modified Snell's law in Eq. \eqref{eq:Klein_aniso_eqs_Snell} and the unique three-surface dispersion in Figs. \ref{Fig_3}(b)-(d).
This new phenomenon can be denoted by the omnidirectional Klein-like tunneling.
Due to the general effective medium formalism in Eq. \eqref{eq:Medium_eq}, this strategy offers a platform for omnidirectional unimpeded wave transmission in diverse classical systems.


\section*{Acknowledgement}

I thank Yair Shokef, Yoav Lahini, Roni Ilan and Moshe Goldstein for useful discussions.



\renewcommand{\thefigure}{S\arabic{figure}}
\renewcommand{\theequation}{S\arabic{equation}}
\setcounter{figure}{0}
\setcounter{equation}{0}

\section*{Appendix A: Acoustic effective medium analogue of Klein tunneling.}  \label{A}


The explicit form of the time-harmonic constitutive relations, Eqs. (2a) and (2b) in domain $1$ and $2$, is given by
\begin{equation}  \label{eq:Medium_eq_exp_1}
\begin{cases} \dfrac{\partial P_1(x,y)}{\partial x} =i\omega m_0V_{1x}(x,y) \\
\dfrac{\partial P_1(x,y)}{\partial y} =i\omega m_0V_{1y}(x,y)  \\
i\omega P_1(x,y)=b_0 \left[\dfrac{\partial V_{1x}(x,y)}{\partial x}+\dfrac{\partial V_{1y}(x,y)}{\partial y}\right] \end{cases} 
\end{equation}
and
\begin{equation}  \label{eq:Medium_eq_exp_2}
    \begin{cases} \dfrac{\partial P_2(x,y)}{\partial x} =i\omega m_0\widetilde{M}(\omega)V_{2x}(x,y) \\
\dfrac{\partial P_2(x,y)}{\partial y} =i\omega m_0\widetilde{M}(\omega)V_{2y}(x,y)  \\
i\omega P_2(x,y)=b_0 \widetilde{B}(\omega) \left[\dfrac{\partial V_{2x}(x,y)}{\partial x}+\dfrac{\partial V_{2y}(x,y)}{\partial y}\right], \end{cases}
\end{equation}
leading to the total wave equations
\begin{equation}   \label{eq:WE_1}
    \frac{\partial^2 P_1(x,y)}{\partial x^2}+\frac{\partial^2 P_1(x,y)}{\partial y^2}=-\frac{\omega^2}{c^2}P_1(x,y)
 \end{equation}
 and
 \begin{equation}   \label{eq:WE_2}
     \frac{\partial^2 P_2(x,y)}{\partial x^2}+\frac{\partial^2 P_2(x,y)}{\partial y^2}=-\frac{\omega^2}{c^2\widetilde{B}(\omega)\widetilde{M}^{-1}(\omega)}P_2(x,y).
 \end{equation}
Substituting traveling wave solutions $P_1(x,y)\propto e^{i(k_{1x}x+k_{1y}y)}$ and $P_2(x,y)\propto e^{i(k_{2x}x+k_{2y}y)}$ for the corresponding pressure fields in Eqs. \eqref{eq:WE_1} and \eqref{eq:WE_2}, gives
\begin{equation}  \label{eq:k_12}
    c^2(k_{1x}^2+k_{1y}^2)=\omega^2 \quad, \quad c^2\widetilde{B}(\omega)\widetilde{M}^{-1}(\omega)(k_{2x}^2+k_{2y}^2)=\omega^2,
\end{equation}
so that the total wavenumbers become
\begin{equation}  \label{eq:k_12_tot}
     k_1=\frac{\omega}{c} \quad, \quad  k_2=k_1\left(\widetilde{M}(\omega)\widetilde{B}^{-1}(\omega)\right)^{1/2}.
\end{equation}
Substituting Eq. \eqref{eq:k_12_tot} into the horizontal stratification condition $k_{1y}=k_{2y}$, i.e. in $k_1\sin\phi=k_2\sin\theta$, results in Eq. (3a).
The first part of Eq. (3b), $1+R=T$, does not depend on the constitutive parameters in domain $2$. It is the direct result of continuity of pressure at $x=0$, $P_1(x=0,y)=P_2(x=0,y)$, or $P_I(x=0,y)+P_R(x=0,y)=P_T(x=0,y)$, where $P_I$, $P_R$ and $P_T$ are the incident, reflected and transmitted fields, explicitly defined as $P_I(x,y)=P_0e^{i(k_{1x}x+k_{1y}y)}$, $P_R(x,y)=P_0e^{i(-k_{1x}x+k_{1y}y)}$ and $P_T(x,y)=P_0e^{i(k_{2x}x+k_{2y}y)}$. 
The second part of Eq. (3b) does depend on $\widetilde{M}(\omega)$ and $\widetilde{B}(\omega)$.
The requirement on continuity of normal flow velocity, $V_{1x}(0,y)=V_{2x}(0,y)$, or $V_{Ix}(0,y)+V_{Rx}(0,y)=V_{Tx}(0,y)$, by Eqs. \eqref{eq:Medium_eq_exp_1} and \eqref{eq:Medium_eq_exp_2}, implies
\begin{equation}  \label{eq:v_cont_exp}
    \frac{\partial P_I(x,y)}{\partial x}_{x=0}+\frac{\partial P_R(x,y)}{\partial x}_{x=0}=\frac{1}{\widetilde{M}(\omega)}\frac{\partial P_T(x,y)}{\partial x}_{x=0}.
\end{equation}
Differentiating $P_I$, $P_R$ and $P_T$, and using $k_{1y}=k_{2y}$, $k_{1x}=k_1\cos\phi$ and $k_{2x}=k_2\cos\theta$ in Eq. \eqref{eq:v_cont_exp}, gives
\begin{equation}  \label{eq:v_cont_subs}
    (1-R)k_1\cos\phi=\widetilde{M}^{-1}(\omega)Tk_2\cos\theta.
\end{equation}
Combining Eq. \eqref{eq:v_cont_subs} with Eq. \eqref{eq:k_12}, Eq. (3b) is retrieved.  

\section*{Appendix B: Acoustic metamaterial realization of the constitutive parameters.}   \label{B}

\begin{figure}[htpb]
    \centering
    \includegraphics[width=5.4 cm, valign=c]{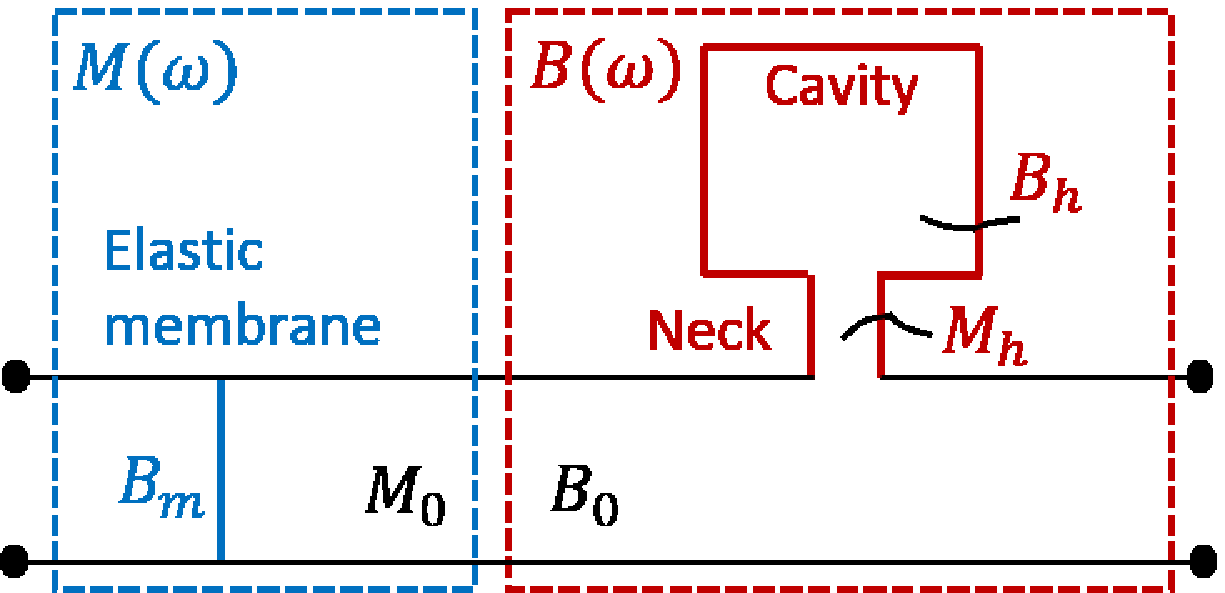}
    \caption{Metamaterial unit cell (closed resonator) in one direction.}
    \label{fig:unit_cell}
\end{figure}

The physics of an acoustic cavity-on-neck resonator, aka Helmholtz resonator, as well as sound wave transmission through an elastic membrane, is well-known \cite{seo2012acoustic}. However, their collective dynamic behavior in the metamaterial setting, producing Eq. (5), requires some derivation.
The derivation here includes dissipation that naturally exists in both membranes and cavities.
To this end, the schematic of Fig. \ref{fig:unit_cell} is considered, which represents a unit cell of length $a$ in a channel of the metamaterial in Fig. 2(a). This channel has a cross-sectional area $A_c=ad$. The resonator, here closed, can be regarded as an air mass per unit area $M_h$ $[kg/m^2]$ attached to an air spring per unit area $B_h$ $[N/m^3]$, with dissipation $D_h$, where the neck of area $A_n=\pi r^2$ stands for the mass, the cavity of volume $Vol$ for the spring, and both are given by
\begin{equation}  \label{eq:Mh_B_h}
    M_h=m_0l\frac{A_c}{A_n} \quad , \quad B_h=\frac{m_0c^2\pi r^2}{Vol}.
\end{equation}
The connection of the resonator to the tube can be thus represented by a serial connection of a dynamic impedance $z_h(\omega)=M_hi\omega+D_h+B_h/i\omega$ with the air impedance $B_0/i\omega$, $B_0=b_0/a$, leading to the effective bulk modulus of
\begin{equation}   \label{eq:B_s}
    \widetilde{B}(\omega)=\frac{i\omega z_h(i\omega)}{B_0+i\omega z_h(i\omega)}=\frac{\omega_h^2-\omega^2+D_hi\omega}{\omega_b^2-\omega^2+D_hi\omega}.
\end{equation}
Here, $\omega_h^2=B_h/M_h$ and $\omega_b^2=\omega_h^2(1+B_0/B_h)=(B_h+B_0)/M_h$. For $\widetilde{B}(s)$ in Eq. \eqref{eq:B_s} to retrieve $\widetilde{B}(\omega)$ in Eq. (5) in the time-harmonic regime (for small dissipation), $\omega_h^2$ needs to equal zero. This implies that either $M_h\rightarrow \infty$ or $B_h\rightarrow 0$. The latter may be achieved with $Vol\rightarrow \infty$, implying an infinite cavity, or, equivalently, an open neck without a cavity, thus keeping the neck radius $r$ and length $l$ finite. This condition retrieves the relation $\omega_b^2=B_0/M_h=\pi r^2c^2/(a^2dl)$, which equals the potential analogue $V_0$.
As for the elastic membrane, it can be regarded as a dynamic air impedance $z_m(\omega)=M_0i\omega+D_m+B_m/(A_ti\omega)$ of an air mass $M_0=m_0a$ and a spring of stiffness $B_m$ $[N/m]$ per channel area $A_t$, implying an effective mass of
\begin{equation}   \label{eq:M_s}
    \widetilde{M}(\omega)=\frac{z_m(\omega)}{M_0i\omega}=\frac{\omega_m^2-\omega^2+D_mi\omega}{\omega^2}.
\end{equation}
For small dissipation, Eq. \eqref{eq:M_s} retrieves Eq. (5).
The characteristic frequency of the membrane is therefore given by $\omega_m^2=B_m/(M_0A_t)=B_m/(m_0a^2d)$. Equating $\omega_b^2$ with $\omega_m^2$, yields $\pi r^2c^2/(a^2dl)=V_0$, as expected. 

\section*{Appendix C: Omnidirectional Klein-like tunneling by anisotropic design.}  \label{C}

\begin{figure}[htpb] 
\begin{center}
  \begin{tabular}{l l}
  \textbf{a}  & \includegraphics[height=4.8 cm, valign=t]{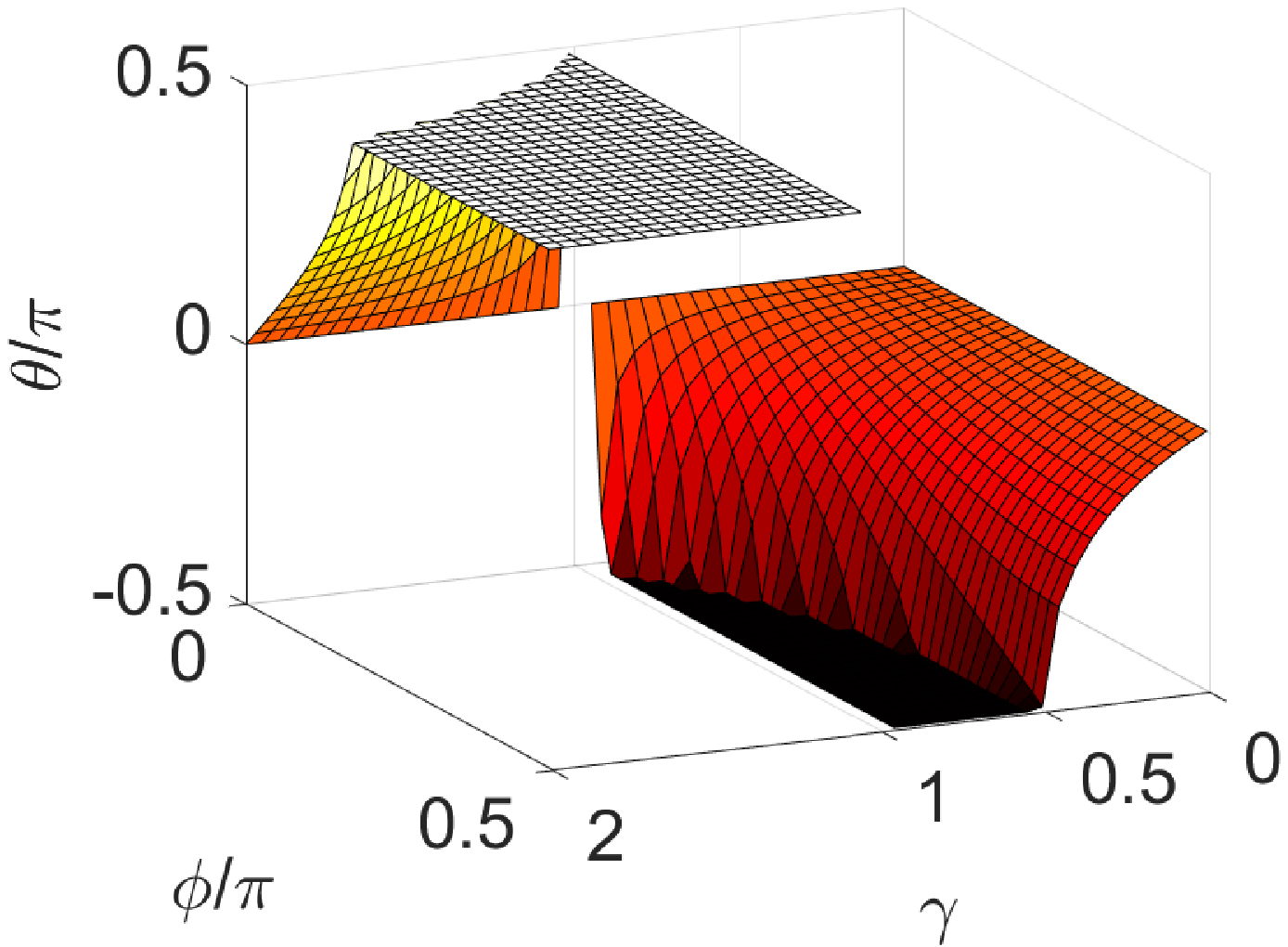} \\ \textbf{b} & \includegraphics[height=4.8 cm, valign=t]{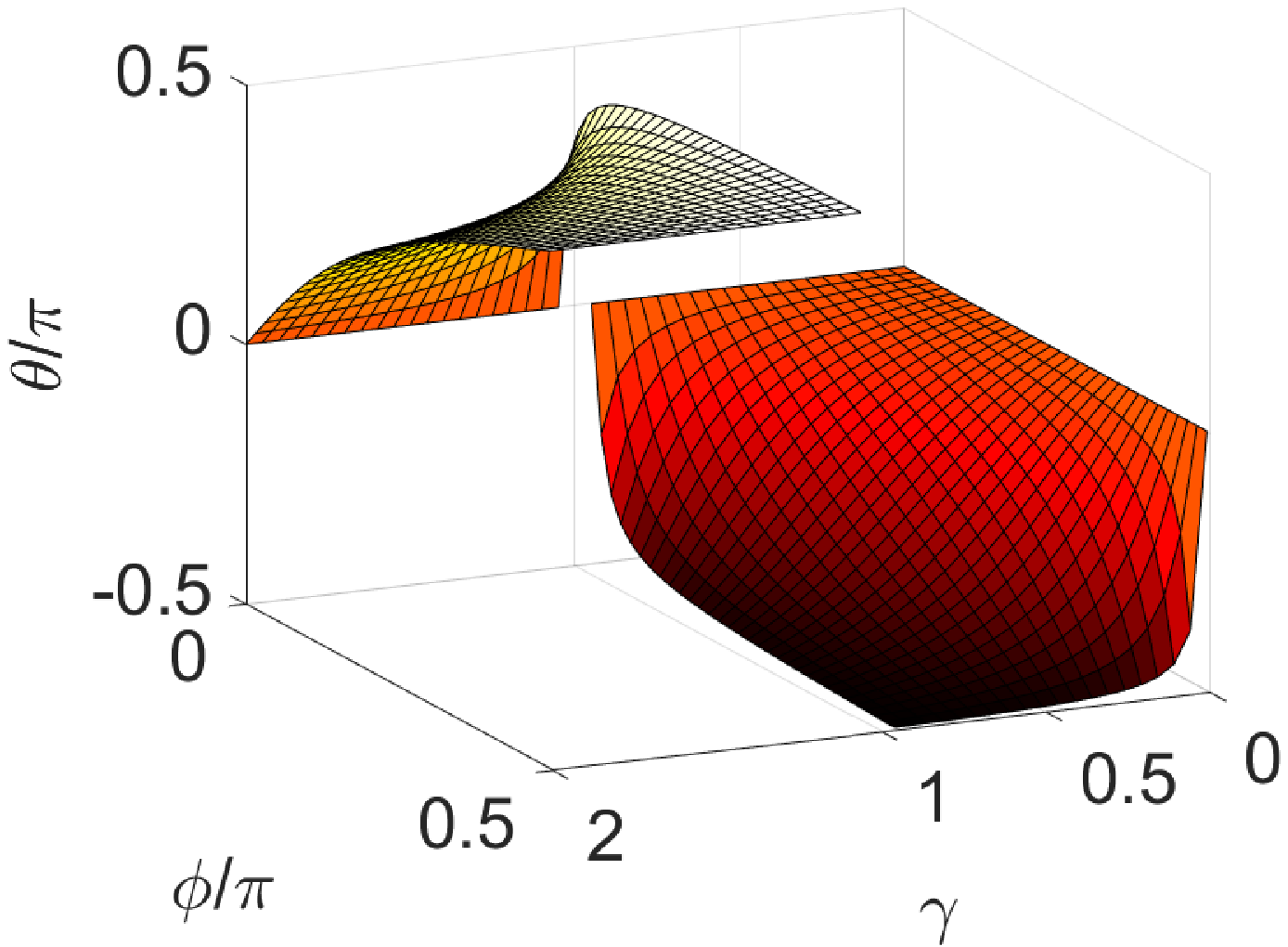}
  \end{tabular}
  \end{center}
   \caption{Refraction angle $\theta$ as a function of $\gamma=\omega^2/V_0$ and incidence angle $\phi$. (a) The isotropic law of refraction in Eqs. \eqref{eq:Klein_like_eqs_Snell}-\eqref{eq:Klein_like_eqs_Fresnel}, with flat surfaces demonstrating existence of critical angle. (b) The modified law in Eqs. \eqref{eq:Klein_aniso_eqs_Snell}-\eqref{eq:Klein_aniso_eqs_Fresnel}, demonstrating no critical angle for any $\gamma$ and $\phi$.}
\label{theta}
\end{figure}

The time-harmonic constitutive equations of the anisotropic medium in domain $2$ take the form
\begin{equation}  \label{eq:Medium_eq_exp_aniso}
\begin{cases} \dfrac{\partial P_2(x,y)}{\partial x} =i\omega m_0\widetilde{M}_x(\omega)V_{2x}(x,y) \\
\dfrac{\partial P_2(x,y)}{\partial y} =i\omega m_0\widetilde{M}_y(\omega)V_{2y}(x,y)  \\
i\omega P_2(x,y)=b_0 \widetilde{B}(\omega) \left[\dfrac{\partial V_{2x}(x,y)}{\partial x}+\dfrac{\partial V_{2y}(x,y)}{\partial y}\right], \end{cases}
\end{equation}
and the total wave equation becomes
\begin{equation}   \label{eq:WE_aniso}
     \widetilde{M}^{-1}_x(\omega)\dfrac{\partial^2 P_2(x,y)}{\partial x^2}+\widetilde{M}^{-1}_y(\omega)\dfrac{\partial^2 P_2(x,y)}{\partial y^2}=\frac{-\omega^2}{c^2\widetilde{B}(\omega)}P_2(x,y).
\end{equation}
With $P_2(x,y)\propto e^{i(k_{2x}x+k_{2y}y)}$, Eq. \eqref{eq:WE_aniso} yields the dispersion relation
\begin{equation} \label{eq:dispersion_aniso}
   c^2\widetilde{B}(\omega)\left(\widetilde{M}^{-1}_x(\omega)k_{2x}^2+\widetilde{M}^{-1}_y(\omega)k_{2y}^2\right)=\omega^2.
\end{equation}
Continuity of normal flow velocity at $x=0$, $V_{1x}(0,y)=V_{2x}(0,y)$, results in
\begin{equation}  \label{eq:v_cont_subs_aniso}
    (1-R)k_{1x}=\widetilde{M}^{-1}_x(\omega)Tk_{2x}.
\end{equation}
For angle-independent perfect transmission it is then required to set
\begin{equation} \label{eq:k_2x}
    k_{2x}=k_{1x}\widetilde{M}_x(\omega),
\end{equation}
which implies $1-R=T$. Together with continuity of pressure requirement, which does not change in the anisotropic regime and yields $1+R=T$, Eq. \eqref{eq:Klein_aniso_eqs_Fresnel} is retrieved. Now, substituting Eq. \eqref{eq:k_2x} into the dispersion relation Eq. \eqref{eq:dispersion_aniso}, and using Snell's law $k_{2y}=k_{1y}$, gives
\begin{equation}
    c^2k_{1x}^2\widetilde{B}(\omega)\widetilde{M}_x(\omega)+c^2k_{1y}^2\widetilde{B}(\omega)\widetilde{M}^{-1}_y(\omega)=\omega^2.
\end{equation}
This needs to retrieve the uniform dispersion relation in domain $1$, which leads to the conditions
\begin{equation}  \label{eq:M_xM_y}
    \widetilde{M}_x(\omega)\widetilde{B}(\omega)=1 \quad , \quad \widetilde{B}(\omega)\widetilde{M}^{-1}_y(\omega)=1.
\end{equation}
On the other hand, using the explicit form of $k_{2y}=k_{1y}$,
\begin{equation}
    \omega^2\sin^2\phi=c^2k_2^2\sin^2\theta=\left[c^2k_{2x}^2+\omega^2\sin^2\phi\right]\sin^2\theta.
\end{equation}
Solving for $k_{2x}$,
\begin{equation}  \label{eq:k_2x_tan}
    c^2k_{2x}^2\sin^2\theta=\omega^2\sin^2\phi(1-\sin^2\theta)=\omega^2\sin^2\phi\cos^2\theta,
\end{equation}
gives
\begin{equation}
    c^2k_{2x}^2\tan^2\theta=\omega^2\sin^2\phi,
\end{equation}
which together with Eq. \eqref{eq:k_2x} and $k_{1x}=\omega/c\cos\phi$, results in
\begin{equation}  \label{eq:theta_phi}
    \tan\phi=\widetilde{M}_x(\omega)\tan\theta.
\end{equation}
Substituting $\widetilde{M}_x(\omega)=\widetilde{M}(\omega)=(\omega^2-V_0)/\omega^2$, Eq. \eqref{eq:theta_phi} retrieves Eq. \eqref{eq:Klein_aniso_eqs_Snell}.
As illustrated in Fig.\ref{theta}\textbf{a}, the critical angle in the isotropic refraction law in Eqs. \eqref{eq:Klein_like_eqs_Snell}-\eqref{eq:Klein_like_eqs_Fresnel} is manifested by flattening of the $\theta$ surface for $\frac{1}{2}<\gamma$, which does not exist in the modified law in Eqs. \eqref{eq:Klein_aniso_eqs_Snell}-\eqref{eq:Klein_aniso_eqs_Fresnel}, Fig.\ref{theta} \textbf{b}.

With $\widetilde{M}_x(\omega)=\widetilde{M}(\omega)$, $\widetilde{B}(\omega)=\widetilde{M}^{-1}_x(\omega)$, $\widetilde{M}_y(\omega)=(\omega^2-\alpha)/\omega^2$, and $\alpha=V_0\gamma_0/(1-\gamma_0)$, as defined in Eq. \eqref{eq:My_aniso}, the dispersion relation Eq. \eqref{eq:dispersion_aniso} takes the form
\begin{equation} \label{eq:aniso_disp_explicit}
\begin{split}
    \gamma^3&-\left(\frac{\alpha}{V_0}+2+\frac{c^2}{V_0}\left(k_{2x}^2+k_{2y}^2\right)\right)\gamma^2\\&+\left(1+2\frac{\alpha}{V_0}+\frac{c^2}{V_0}\left(\frac{\alpha}{V_0}k_{2x}^2+k_{2y}^2\right)\right)\gamma-\frac{\alpha}{V_0}=0,
    \end{split}
\end{equation}
where $\gamma=\omega^2/V_0$ for the general frequency $\omega$, and $\gamma_0=\omega_0^2/V_0$ for the specific working frequency $\omega_0$. The relation in Eq. \eqref{eq:aniso_disp_explicit} represents the dispersion plots in Figs. 4(b)-(d). 

\begin{figure}[htpb]
\begin{center}
\setlength{\tabcolsep}{0pt}
    \begin{tabular}{c c c}
     &  $\gamma_0=\frac{2}{3}$ & $\gamma_0=\frac{1}{3}$ \\
     \rotatebox{90}{Top surface} & \includegraphics[height=\x cm, valign=c]{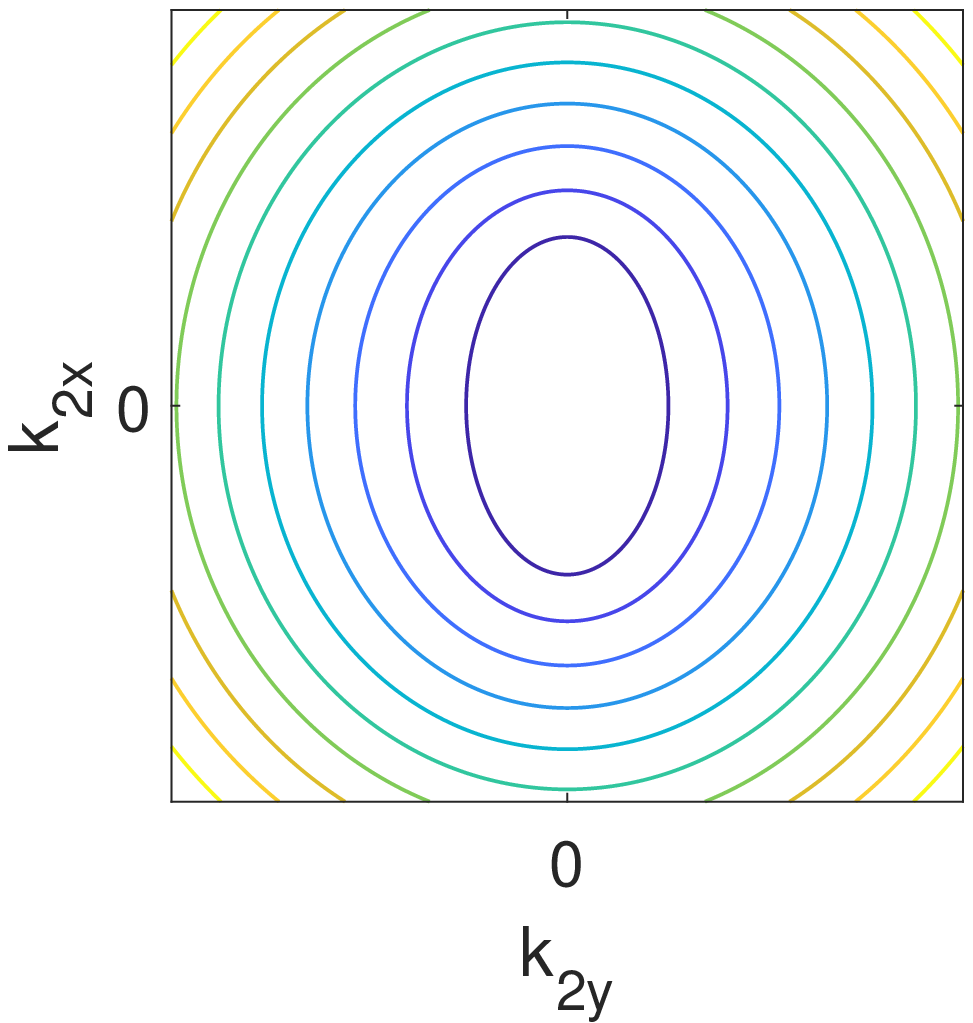}  & \includegraphics[height=\x cm, valign=c]{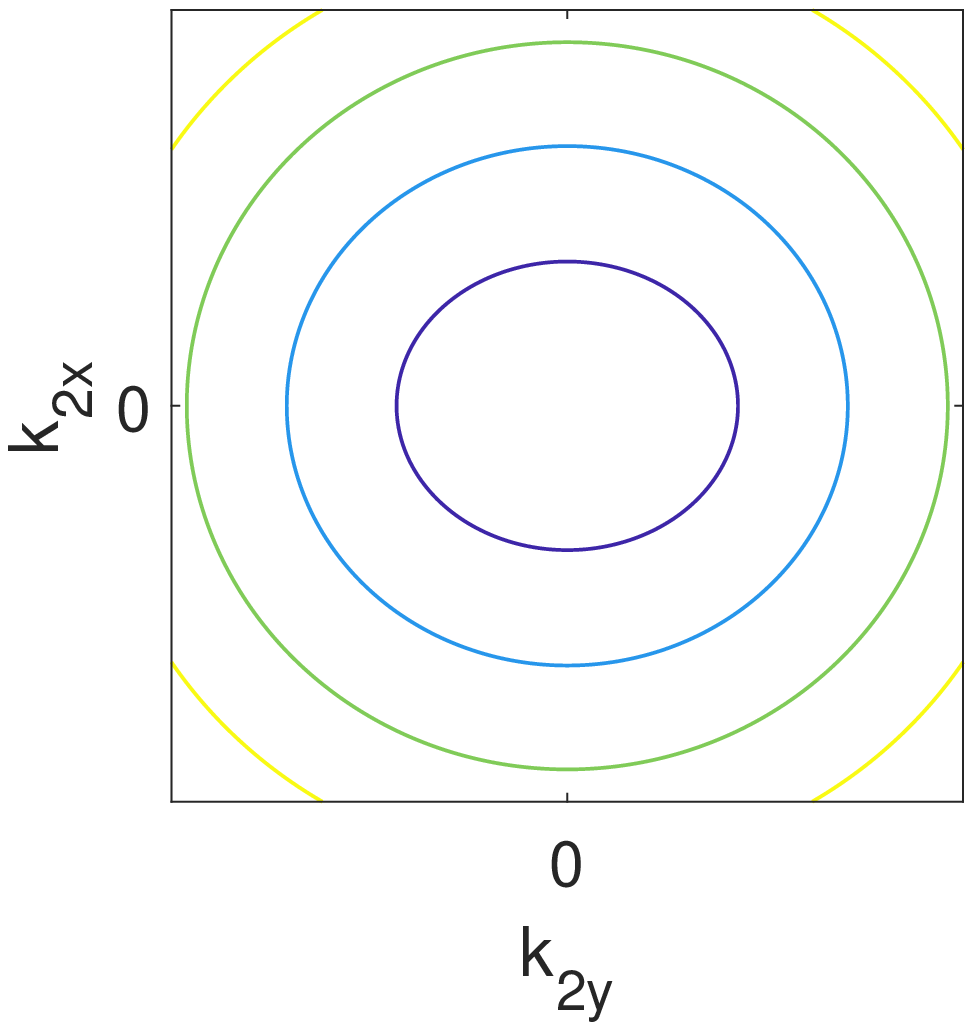} \\
     \rotatebox{90}{Middle surface} & \includegraphics[height=\x cm, valign=c]{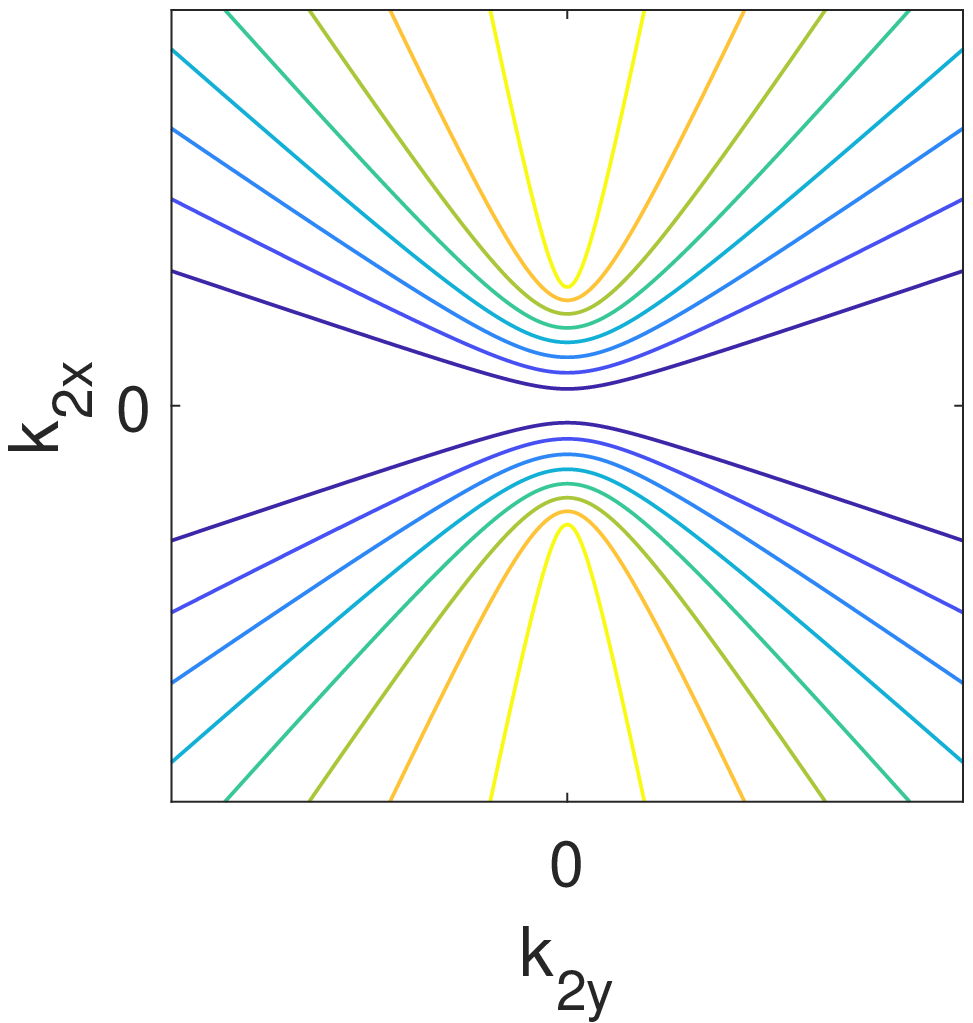} &  \includegraphics[height=\x cm, valign=c]{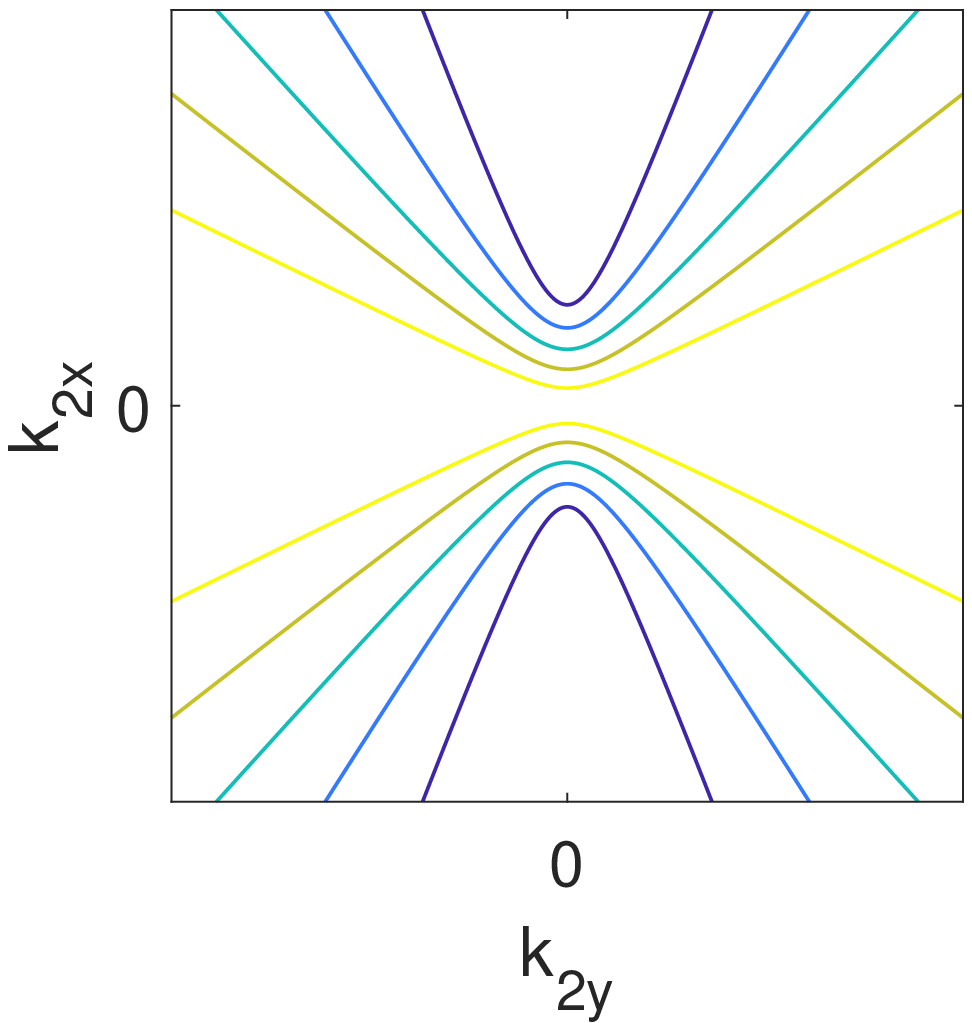} \\
     \rotatebox{90}{Bottom surface} & \includegraphics[height=\x cm, valign=c]{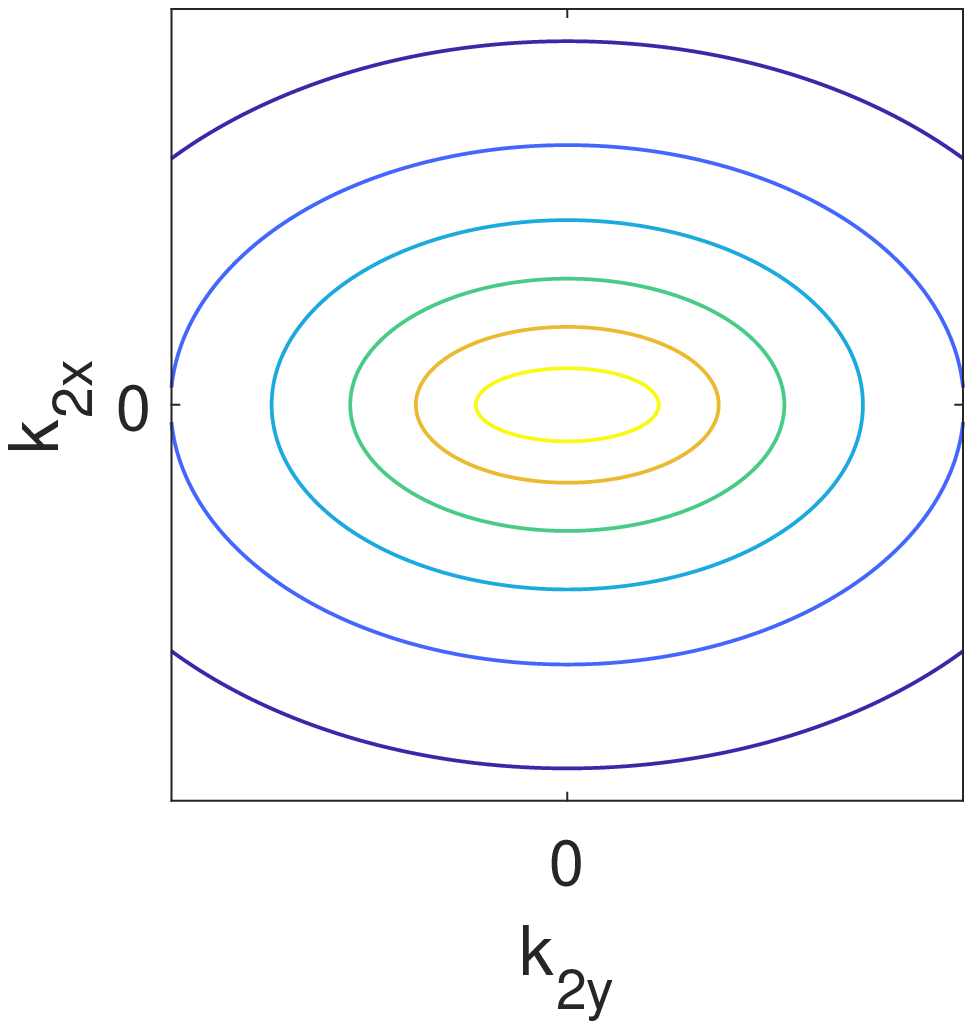}  & \includegraphics[height=\x cm, valign=c]{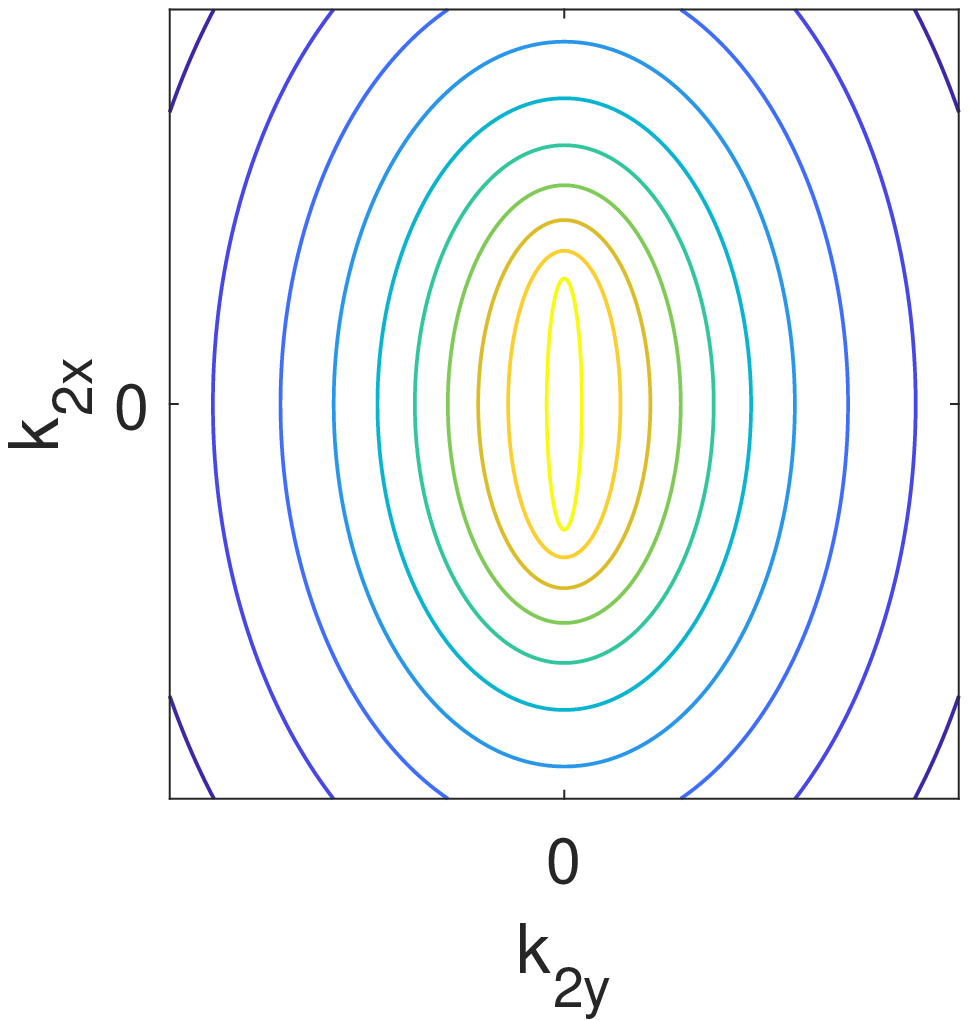}
    \end{tabular}
\end{center}
\caption{Isofrequency contours of the omnidirectional Klein-like tunneling dispersion plots for $\gamma_0=2/3$ and $1/3$. Top and bottom surfaces are of an elliptic cross-section, whereas the middle surface is of a hyperbolic cross-section. Polarization flipping occurs in the transition through $\gamma_0=1/2$.}
\label{contour}
\end{figure}

%
In Fig. \ref{contour}, isofrequency contours of the dispersion surfaces of Figs. 4(b)-(d) are depicted. 
For $1/2<\gamma_0<1$ and $0<\gamma_0<1/2$, the isofrequency cross-sections of the middle surface are hyperbolic, forming tilted Dirac-like cones with the top or bottom surfaces, which have elliptic cross-sections.
%
At $\gamma_0=1/2$ the middle surface degenerates, and the top and bottom surfaces become regular circular cones, touching at the origin. 
The elliptic cones polarization is flipped between $1/2<\gamma_0<1$ and $0<\gamma_0<1/2$, with the bottom cone indicating the interplay of the tunnelled wave group and phase velocity directions.

\begin{figure}[tb] 
\begin{center}
  \begin{tabular}{c c c}
$\gamma_0$ & $\phi=15^o$ & $\phi=35^o$ \\
$\frac{2}{3}$ & \includegraphics[height=4.2 cm, valign=c]{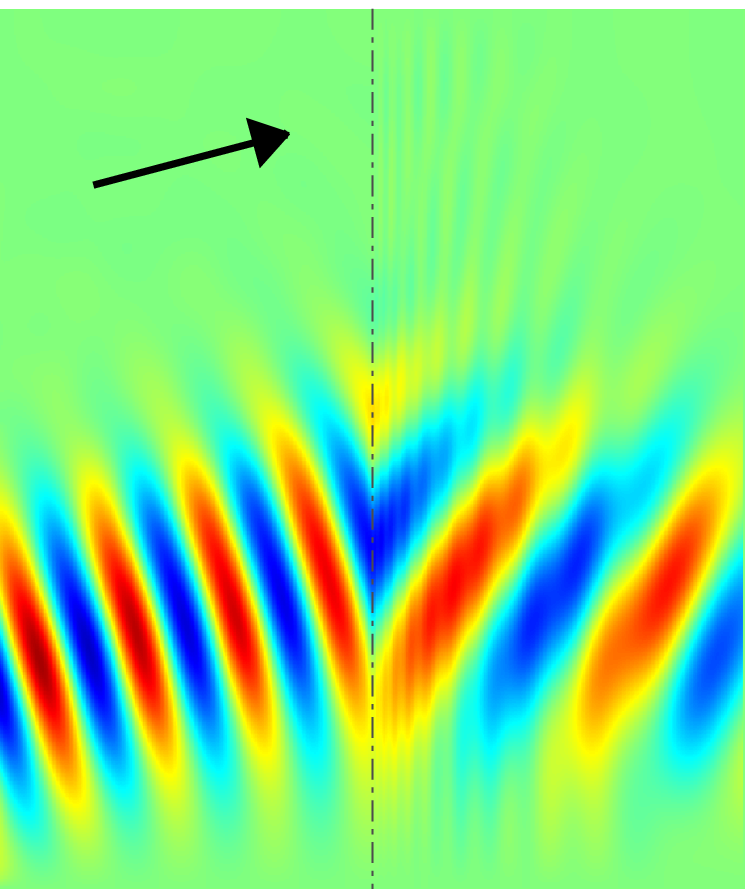} & \includegraphics[height=4.2 cm, valign=c]{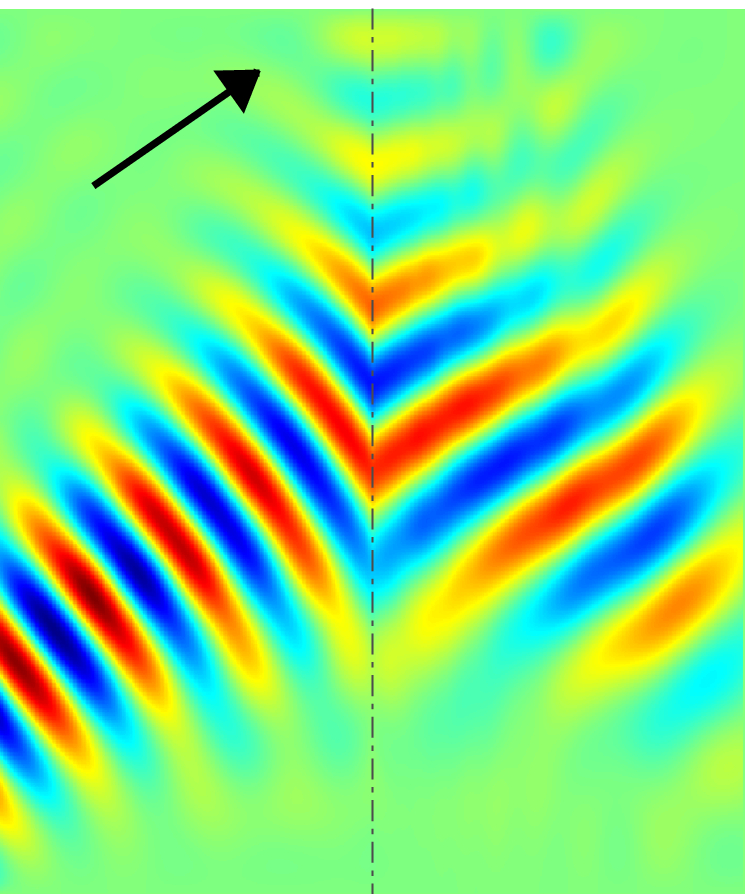} \\
 & & \\
$\frac{1}{2}$ & \includegraphics[height=4.2 cm, valign=c]{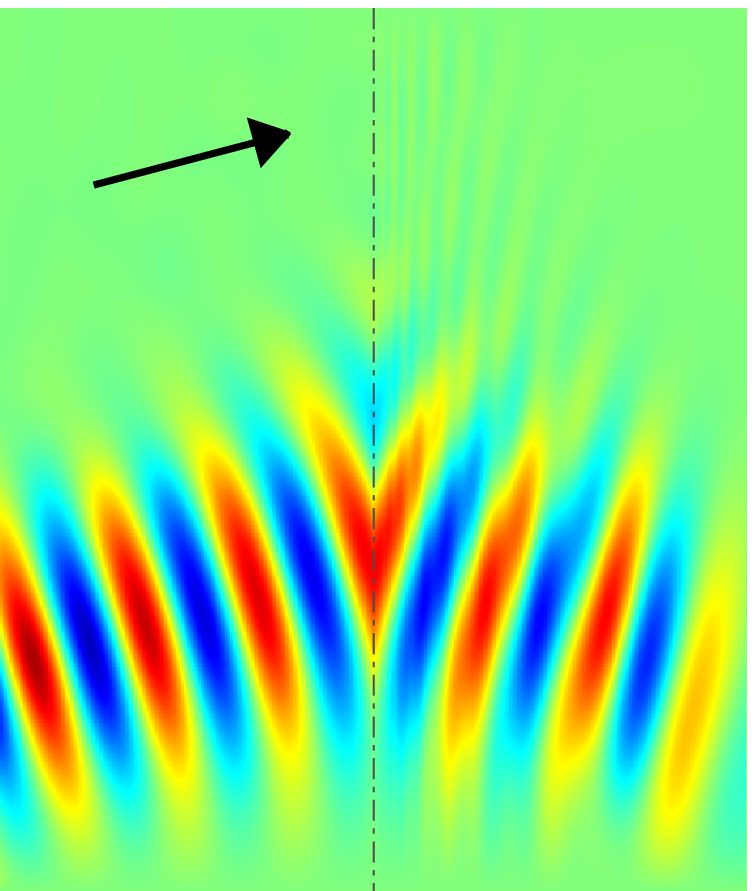} & \includegraphics[height=4.2 cm, valign=c]{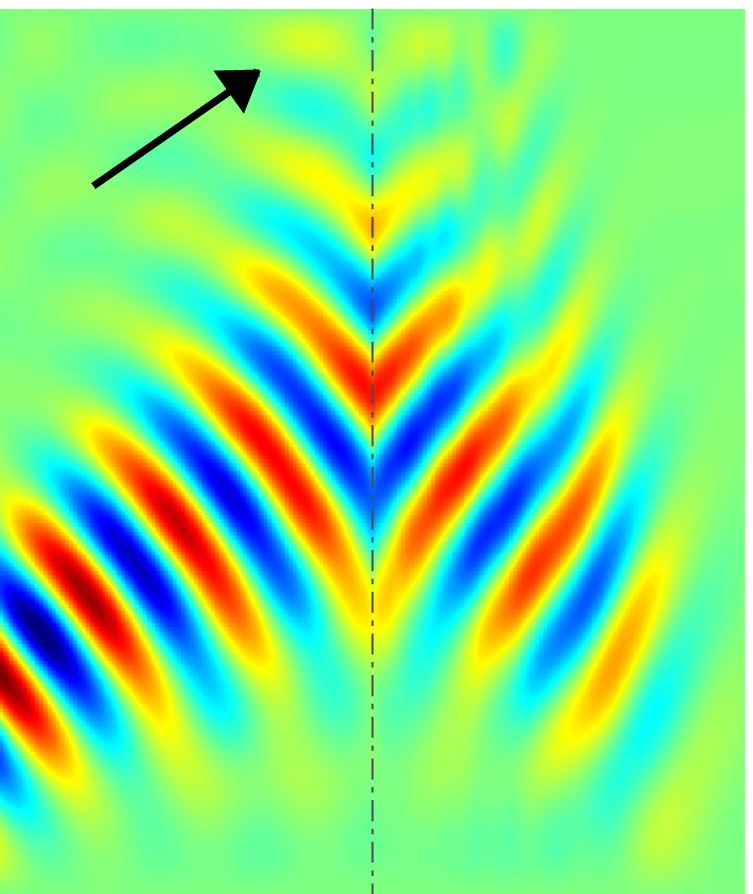} \\
& & \\
$\frac{1}{3}$ & \includegraphics[height=4.2 cm, valign=c]{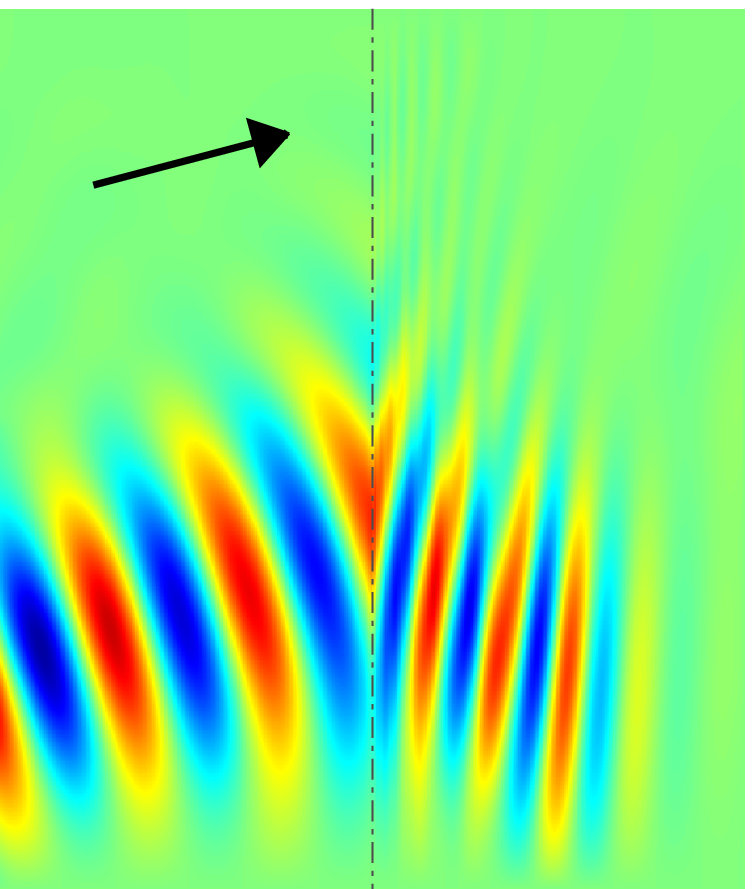} & \includegraphics[height=4.2 cm, valign=c]{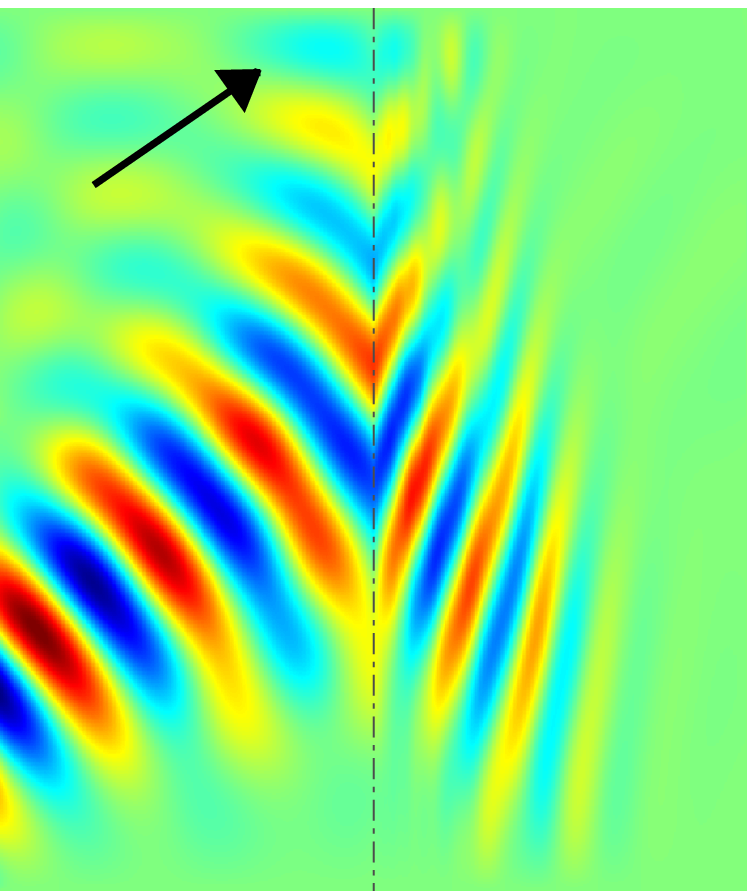} \\ & &
\end{tabular}
\includegraphics[width=3.8 cm]{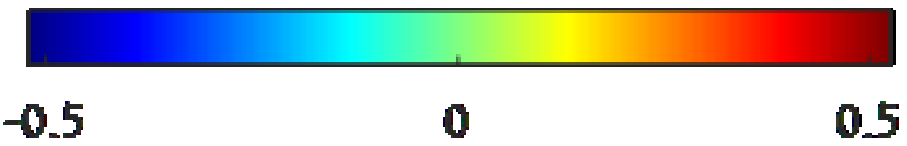} 
\end{center}
 \caption{Demonstrating omnidirectional Klein-like tunneling for two additional incidence angles.}
\label{Fig_4}
\end{figure}

To illustrate the angle independence of the omnidirectional Klein-like tunneling, dynamical simulations of the anisotropic medium for two additional incidence angles, $\phi=15^o$ and $\phi=35^o$, are depicted in Fig. \ref{Fig_4}. 


\begin{figure}[tb] 
\begin{center}
  \begin{tabular}{c c c}
$\gamma_0$ &  medium & metamaterial \\
$\frac{2}{3}$ & \includegraphics[height=4.2 cm, valign=c]{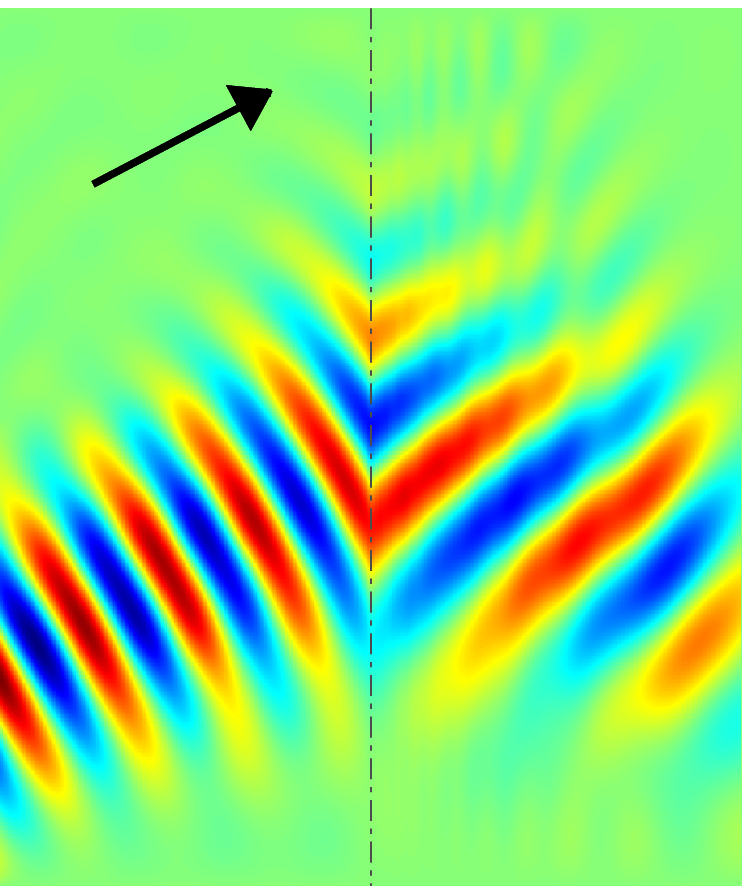} & \includegraphics[height=4.2 cm, valign=c]{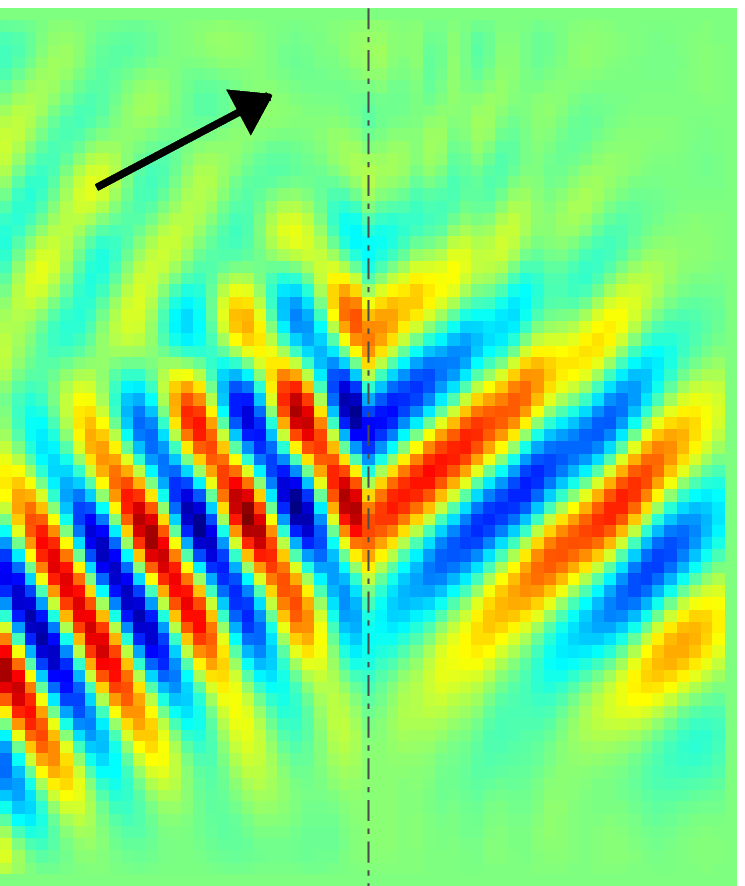} \\
 & & \\
$\frac{1}{2}$ & \includegraphics[height=4.2 cm, valign=c]{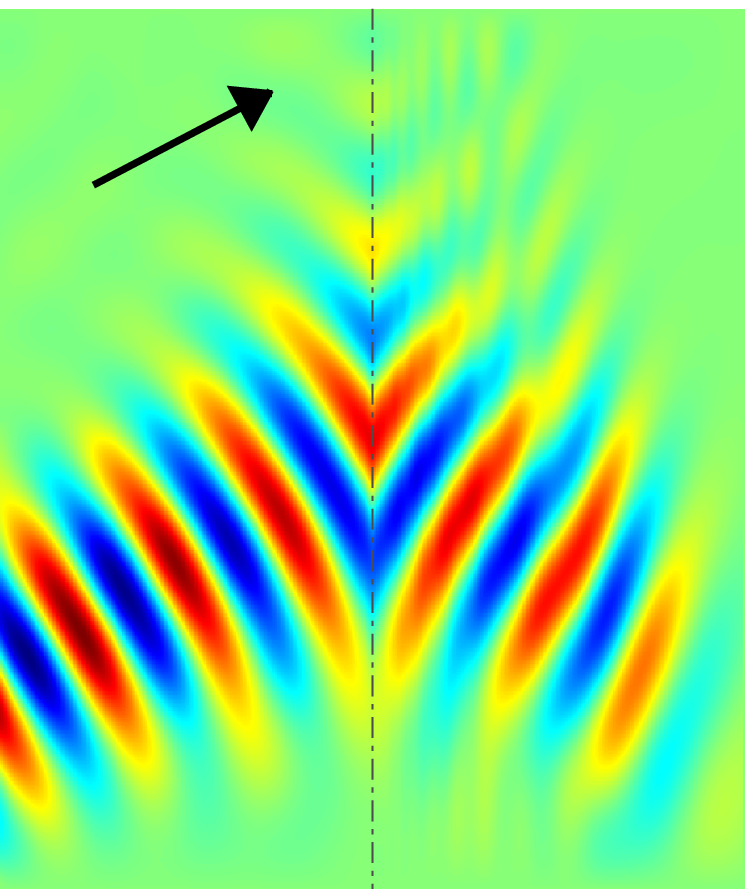} & \includegraphics[height=4.2 cm, valign=c]{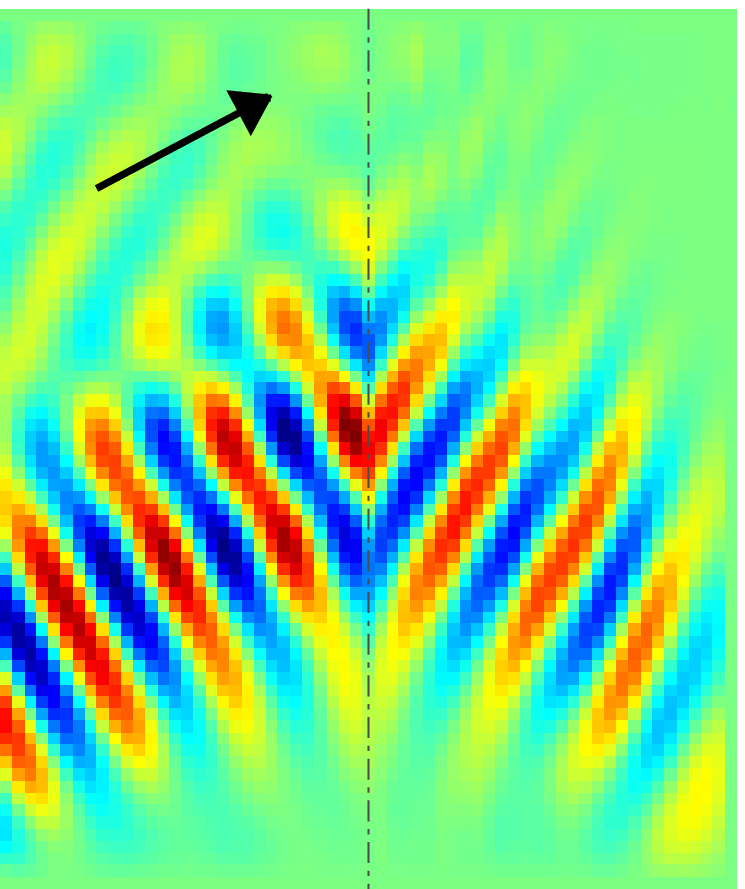} \\
& & \\
$\frac{1}{3}$ & \includegraphics[height=4.2 cm, valign=c]{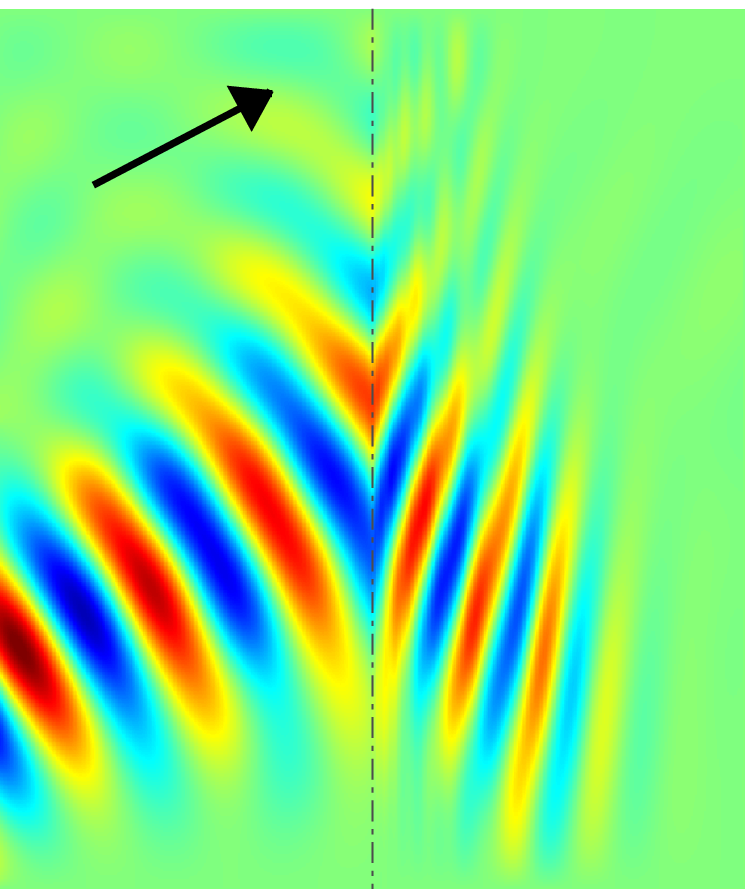} & \includegraphics[height=4.2 cm, valign=c]{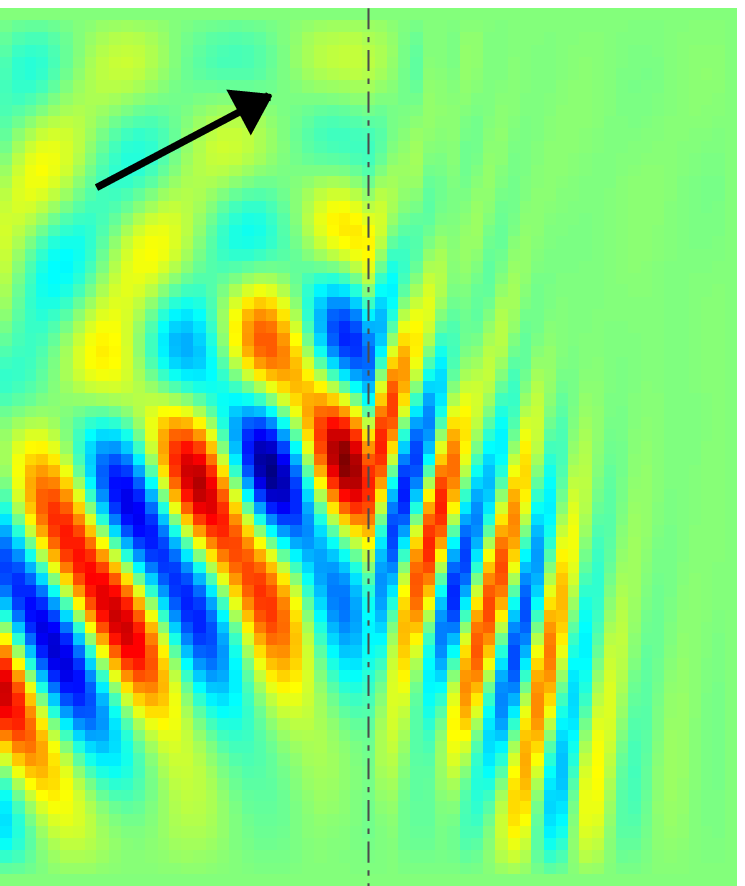} \\ & &
\end{tabular}
\includegraphics[width=3.8 cm]{ISF_Figures/colorbar_jet.eps} 
\end{center}
 \caption{Omnidirectional Klein-like tunneling - comparing continuous medium and actual discrete metamaterial time domain responses.}
\label{Fig_5}
\end{figure}

Finally, the homogenized medium response for the incidence angle of $28^o$ is compared in Fig. \ref{Fig_5} to the response of the actual discrete metamaterial with unit cell size of $a\approx\lambda/8$. 

\section*{Appendix D: Finite difference time domain simulation details.}  \label{D}

The computational algorithm was based on coupled spatial and temporal iterative schemes, where the spatial scheme extending in both $x$ and $y$ directions. The update in time was carried out using a step of $dt=10^{-6}$. The space was discretized by a step of $dx=\lambda/40$ in the continuous medium case, and $dx=\lambda/8$ in the discrete structure case. 

The source function was a continuous sinusoidal wave generated at the left boundary, $x=0$. To localize the wave at a finite section and thus to create a finite width beam, the wave was truncated along the vertical $y$ axis by a Gaussian of width proportional to $\sigma^2$. The tilt angle $\phi$ of the beam was created by inducing a targeted phase shift in the source as a function of the distance $y$, $\Delta\varphi(y)=(2\pi a)/\lambda_1\sin{\phi}$. The overall source function for the discretization step $a$ and incidence domain (medium 1) wavelength $\lambda_1$, was therefore given by
\begin{equation}
    f(y,t)=e^{-\frac{(y\sigma)^2}{2}}\sin{\left(\omega t +\Delta\varphi(y)\right)}.
\end{equation}

For domain truncation the absorbing boundary conditions (ABCs) technique was used, which matches the medium impedance $z_0$ to that of the boundary. In general, this is not trivial when the medium is dispersive, since the effective medium impedance is frequency dependent. In our case, however, the impedance  $z_2(\omega)=z_0$ implied the ABCs
\begin{equation}
    \frac{\partial p(x,y,t)}{\partial t}|_{x=L}=-\frac{z_0}{\rho}\frac{\partial p(x,y,t)}{\partial x}|_{x=L}
\end{equation}
along the $x=L$ boundary. In the $y$ direction an extended domain was used to allow the wave to hit the $x=L$ boundary. The implementation of the frequency dispersion was a three-step process: (i) translating the frequency domain expressions to time domain differentiation operators (e.g. $-\omega^2p\Leftrightarrow \partial ^2p/\partial t^2$), resulting in higher order partial differential equations (PDEs), (ii) converting the PDEs into an auxiliary system of first order equations, and (iii) augmenting the iteration scheme accordingly with the spatially discretized version of the auxiliary equations.

The frequency domain functions are $P(x,y)$, $V_x(x,y)$ and $V_y(x,y)$, whereas the corresponding time domain functions are $p(x,y,t)$, $v_x(x,y,t)$ and $v_y(x,y,t)$. Considering the anisotropic case dispersion defined in Eq. (8), the explicit form of medium 2 equations in frequency and time domain respectively become
    \begin{equation}  \label{eq:Medium_eq_iso}
\begin{cases}
    \dfrac{\partial P}{\partial x}=i\omega m_0\dfrac{\omega^2-V_0}{\omega^2}V_x \\ \dfrac{\partial P}{\partial y}=i\omega m_0\dfrac{\omega^2-\alpha}{\omega^2}V_y    \\ i\omega P=b_0\dfrac{\omega^2}{\omega^2-V_0}\left(\dfrac{\partial V_x}{\partial x}+\dfrac{\partial V_y}{\partial y}\right)
\end{cases}
\end{equation}
and
\begin{equation}
    \begin{cases}
    \dfrac{\partial^2 p}{\partial x \partial t}= -m_0\left[\dfrac{\partial^2 v_x}{\partial t^2}-V_0v_x\right] \\ \dfrac{\partial^2 p}{\partial y \partial t}= -m_0\left[\dfrac{\partial^2 v_y}{\partial t^2}-\alpha v_y\right] \\ \dfrac{\partial^2 p}{\partial t^2}-V_0p=-b_0\left(\dfrac{\partial^2 v_x}{\partial x \partial t}+\dfrac{\partial^2 v_y}{\partial y \partial t}\right).
\end{cases}
\end{equation}
Defining the auxiliary variables
\begin{equation}  \label{eq:iso_def}
    q=\dfrac{\partial p}{\partial t} \quad , \quad a_x=\dfrac{\partial v_x}{\partial t} \quad , \quad a_y=\dfrac{\partial v_y}{\partial t}
\end{equation}
the high order PDEs in \eqref{eq:Medium_eq_iso} can be rewritten in the form
\begin{equation}  \label{eq:iso_aux}
    \begin{cases}
        \dfrac{\partial q}{\partial x}= -m_0\left(\dfrac{\partial a_x}{\partial t}-V_0v_x\right) \\ \dfrac{\partial q}{\partial y}= -m_0\left(\dfrac{\partial a_y}{\partial t}-\alpha v_y\right) \\ \dfrac{\partial q}{\partial t}-V_0p=-b_0\left(\dfrac{\partial a_x}{\partial x}+\dfrac{\partial a_y}{\partial y }\right).
    \end{cases}
\end{equation}
The system of the first order PDEs in \eqref{eq:iso_def}-\eqref{eq:iso_aux}, once spatially discretized and rearranged for the update in time, constitutes the finite difference time domain scheme used in the simulations of Fig. 4(e)-(g). The simulations of the isotropic case, Fig. 2 (b),(c), were carried out using the same scheme, by substituting $\alpha=V_0$.

Two types of dissipation were modeled. One was an overall dissipation $\zeta$, which affects the entire pressure field as $\zeta \frac{\partial p}{\partial t}$, and exists in both in medium 1 and 2. The other type was a dissipation $\zeta_m$ caused by particular features of the metamaterial cells in medium 2, thus directly affecting the effective constitutive parameters. The latter, with equal values of 0.005 for brevity, was therefore modeled as
\begin{equation}
\begin{cases}
    \widetilde{M}_x(\omega)=\dfrac{\omega^2-i\omega\zeta_m-V_0}{\omega^2} \; , \;
    \widetilde{M}_y(\omega)=\dfrac{\omega^2-i\omega\zeta_m-\alpha}{\omega^2} \\
    \widetilde{B}(\omega)=\dfrac{\omega^2}{\omega^2-i\omega\zeta_m-V_0}.
\end{cases}
\end{equation}


\bibliographystyle{IEEEtran}        
\bibliography{AcousticKlein}           

\end{document}